\documentclass[preprint,longbibliography,aps,pra,superscriptaddress,showpacs,floatfix]{revtex4-1}
\usepackage{graphicx}
\usepackage{wrapfig}
\usepackage{dcolumn}
\usepackage{bm}
\usepackage{amsfonts}
\usepackage{amsmath}
\usepackage{cancel}
\usepackage{multirow}
\usepackage{color}
\usepackage{ulem}
\usepackage[utf8]{inputenc}

\newcommand {\be}{\begin{equation}}
\newcommand {\ee}{\end{equation}}
\newcommand {\bea}{\begin{eqnarray}}
\newcommand {\ea}{\end{eqnarray*}}
\newcommand {\ba}{\begin{eqnarray*}}
\newcommand {\eea}{\end{eqnarray}}

\begin{document}

\title{Energy spectra of small bosonic clusters having a large two-body
      scattering length} 

\author{M. Gattobigio}
\affiliation{Universit\'e de Nice-Sophia Antipolis, Institut Non-Lin\'eaire de
Nice,  CNRS,
1361 route des Lucioles, 06560 Valbonne, France }
\author{A. Kievsky}
\author{M. Viviani}
\affiliation{Istituto Nazionale di Fisica Nucleare, Largo Pontecorvo 3, 56100 Pisa, Italy}

\begin{abstract}
In this work we investigate small clusters of bosons using the
hyperspherical harmonic basis. We consider systems with $A=2,3,4,5,6$ particles
interacting through a soft inter-particle potential. In order to make contact 
with a real system, we use an attractive gaussian potential that reproduces 
the values of the dimer binding energy and the atom-atom scattering length 
obtained with one of the most widely used $^4$He-$^4$He interactions, the LM2M2 potential.  
The intensity of the potential is varied in order to explore the clusters' spectra  
in different regions with large positive and large negative values of 
the two-body scattering length. In addition, we include a repulsive three-body 
force to reproduce the trimer binding energy. With this model, consisting in the 
sum of a two- and three-body potential, we have calculated the spectrum of the
four, five and six particle systems. In all the region explored, we have found that 
these systems present two states, one deep and one shallow close to the $A-1$ 
threshold. Some universal relations between the energy levels are extracted; in
particular, we have estimated the universal ratios between thresholds of the 
three-, four-, and five-particle continuum using the two-body gaussian
potential. They agree with recent measurements and theoretical predictions.
\end{abstract}

\maketitle

\section{Introduction}

Systems composed by few atoms having large value of the two-body scattering
length, $a$,
with respect to the natural length, $\ell$, fixed by the atomic potential, 
have been the object of an intense investigation both from a theoretical and experimental 
point of view (for recent reviews see 
Refs.~\cite{braaten:2006_physicsreports,greene:2010_phys.today,ferlaino:2010_physics}).
In fact, 
they present universal properties: for example, the three-body system displays 
the Efimov effect~\cite{efimov:1970_phys.lett.b,efimov:1971_sov.j.nucl.phys.},
that means the appearance, 
in the limit  $a/\ell \rightarrow\infty$,
of an infinite set of bound states accumulating 
toward the three-particle threshold; moreover, the 
three-body spectrum has a discrete-scale symmetry,
with an universal ratio between the  $n$-th and $n+1$-th levels 
$E^{n+1}_3/E^n_3={\rm e}^{-2\pi/s_0}$. 
The scaling factor depends only on the ratio
between particle masses, and for identical bosons of mass $m$ it reads
$\displaystyle\text{e}^{-2\pi/s_0} \approx 1/515.03$ (with $s_0\approx1.00624$).  
The finite value of $\ell$ implies the existence of a three-body ground state
$E_3^0$ whose value reflects the short range physics, and that, together with
the discrete-scale symmetry, completely determines the spectrum.
In realistic cases the ratio $a/\ell$ is large but finite; thus, the
three-body spectrum reduces to a finite number of states. 

A remarkable property in $a \rightarrow\infty$
limit appears in the four-body system: two states 
$E_4^{n,0}, E_4^{n,1}$ are attached to each trimer state $E_3^n$, one deep and 
one shallow having universal ratios, $E_4^{n,0}/E_3^n\approx 4.6$ and 
$E_4^{n,1}/E_3^n\approx
1.001$~\cite{platter:2004_phys.rev.a,von_stecher:2009_natphys,deltuva:2011_few-bodysyst.};
the two lowest four-body states, 
$E_4^0 = E_4^{0,0}$ and  $E_4^1 = E_4^{0,1}$,
are real bound states. These properties have been studied for
large positive and large negative values of the scattering length in the
$(a^{-1},\kappa)$ plane, with $\kappa={\rm sign}(E)[|E|/(\hbar^2/m)]^{1/2}$, constructing
what is normally called an Efimov plot~\cite{hammer:2007_eur.phys.j.a}. 

There are very few studies of the spectrum of small bosonic clusters beyond
$A=4$. In addition to the specific problems related to the solution of the
Schr\"odinger equation for more than four particles, the atom-atom realistic
potentials present a strong repulsion at short distances which makes the
numerical problem more difficult.  Specific algorithms have been developed so
far to solve this problem: the Faddeev equation has been opportunely
modified~\cite{kolganova:1998_j.phys.b},  the Hyperspherical methods resorted
either to the hyperspherical adiabatic (HA) expansion  (for a review see
Ref.~\cite{nielsen:2001_phys.rep.}), or to the correlated hyperspherical
harmonic expansion (CHH)~\cite{barletta:2001_phys.rev.a}.  However, due to the
difficulties in treating the strong repulsion, few calculations exist for
systems with more than three atoms. For example, in
Ref.~\cite{lewerenz:1997_j.chem.phys.} the diffusion Monte Carlo method has been used
to describe the ground state of $^4$He molecules up to 10 atoms, and in
Ref.~\cite{hiyama:2012_phys.rev.a} a very extended calculation has been done in
the four helium atom system. On the other hand, descriptions of few-bosons
systems using soft-core potentials are currently operated (see for example
Refs.~\cite{von_stecher:2009_natphys,timofeyuk:2008_phys.rev.c}).  

The equivalence between hard- or soft-core-potential descriptions
has been discussed in Refs.~\cite{kievsky:2011_few-bodysyst.,gattobigio:2011_phys.rev.a}, 
in which an
attractive soft $^4$He-$^4$He gaussian potential has been used to investigate the
three-atom
system. The soft-two-body potential was designed to reproduce the helium
dimer binding energy $E_{2}$, the $^4$He-$^4$He scattering length $a$, and the
effective range $r_0$ of the LM2M2 potential \cite{aziz:1991_j.chem.phys.}, one
of the most used $^4$He-$^4$He interactions. In this context the soft
gaussian potential can be considered as a regularized-two-body contact term
in an Effective Field Theory (EFT) approximation of the
LM2M2~\cite{lepage:1997_}; this is possible because of the scale separation
between the $^4$He-$^4$He scattering length, $a=189.41$~a.u., and the natural length
$\ell=10.2$~a.u., which is the van der Waals length calculated for the LM2M2
potential~\cite{braaten:2006_physicsreports}.

In the two-body sector and in the low-energy
limit, the two potentials predict similar phase shifts, therefore, even if
their shape is completely different, they describe in an equivalent way the
physical processes in that limit~\cite{lepage:1997_}.  The equivalence is
lost as the energy is increased, when the details of the potential become more
and more important. When the soft interaction is used
in the three-body sector, a new three-body-contact term is required to
reproduce the ground-state-binding energy of the helium trimer given by
the LM2M2 potential. This term is introduced by means of a
gaussian-hypercentral three-body force, whose 
strength is tuned to reproduce the LM2M2 ground state binding energy of
the three-atom system. In Ref.~\cite{kievsky:2011_few-bodysyst.} the quality of this
description has been studied for different ranges of the three-body force by
comparing the binding energy of the excited Efimov state and the low-energy
helium-dimer phase shifts to those obtained with the LM2M2 potential. 
In Ref.~\cite{gattobigio:2011_phys.rev.a} the spectrum of small clusters of helium atoms
has been investigated up to six particles maintaining however fixed the
values of $a$ and $E_{2}$ as given by the LM2M2 potential.

In the present work we extend the analysis of the $A=3-6$ bosonic spectrum to
the $(a^{-1},\kappa)$ plane. We have modified the strength of the LM2M2 potential in
order to cover the region of negative values of $a$ up to $a^0_-$, with this
value indicating the threshold of having a three-body system bound.
We have also increased the
intensity of the interaction in order to extend the analysis to positive
values of $a$ in which the universal character of the system starts to be
questionable, i.e, when the ground-state $E^0_3$ approaches the natural
energy $E_\ell=-\hbar^2/m\ell^2$, which delimits the Efimov window. 

Associate with the different values of $a$ of the
modified LM2M2 potential, we have
constructed a set of attractive gaussian potentials with the strength 
fixed to reproduce the low-energy data of LM2M2. 
Moreover, the modifications of the LM2M2 produce different
values of the $A=3$ ground state energy $E_3^0$; accordingly, we
introduce a soft three-body force devised to reproduce those values along
the $(a^{-1},\kappa)$ plane. Within this model, consisting in the
sum of a two- and three-body potentials, we have calculated the spectrum of the
four, five and six particle systems.

Two different calculations have been performed in the present work.
From one side we have calculated the $A=3$ ground state and excited state,
$E_3^0$ and $E_3^1$, using the LM2M2 potential and its modification,
in order to construct the corresponding Efimov plot.
Since this potential present a strong short-range repulsion we have used
the CHH expansion as discussed
in  Ref.~\cite{barletta:2001_phys.rev.a}. One the other side, 
when using the soft-core potential model in systems with $A\ge 3$,
the numerical calculations were performed by means of the non-symmetrized 
hyperspherical harmonic 
(NSHH) expansion method with the technique recently developed by the authors in
Refs.~\cite{gattobigio:2009_phys.rev.a,gattobigio:2009_few-bodysyst., gattobigio:2011_phys.rev.c,gattobigio:2011_phys.rev.a}. 
In this approach, the authors have used the Hyperspherical Harmonic (HH) basis, 
without a previous symmetrization procedure, to describe
bound states in systems up to six particles. The method is based on a
particular representation of the Hamiltonian matrix, as a sum of products of
sparse matrices, well suited for a numerical implementation. Converged results
for different eigenvalues, with the corresponding eigenvectors belonging to
different symmetries, have been obtained~\cite{gattobigio:2011_phys.rev.c}. 
In the present work, since we are dealing 
with bosons, we only consider the symmetric part of the spectrum. 
Interestingly, we have observed that in all the region explored the 
$A=4,5,6$ systems present two states, one deep and one shallow close to the
$E_{A-1}^0$ threshold.
To gain insight on the shallow state, for a selected value of $a$, we have
varied the range of the three-body force and we have studied the effect of that
variation in the $A=4,5,6$ spectrum. In the range considered, the variation
produces small changes in the eigenvalues, but they are crucial to determine if
the shallow state is bound or not with respect to the $A-1$ threshold. This
analysis confirms, at least in one zone of the Efimov plot, previous
observations that each Efimov state in the $A=3$ system produces two bound
states in the $A=4$ system, and extends this observation to the $A=5,6$ systems.

Finally, we have extended the calculations of the $A=4$ and $A=5$ systems up to
the four- and five-particle thresholds using the simple two-body-gaussian
potential; the ratios between the thresholds are in agreement with previous
theoretical
results~\cite{von_stecher:2009_natphys,von_stecher:2010_j.phys.b:at.mol.opt.phys.}
and with 
experiments~\cite{pollack:2009_science,zaccanti:2009_natphys,ferlaino:2009_phys.rev.lett.,zenesini:2012_}.

The paper is organized as follows. In Section II we describe the two-
and three-body forces we used in our calculations to reproduce the LM2M2 
values. In Section III we discuss the Efimov plot for three particles.
In Section IV the results for the bound states of the $A=3,4,5,6$ clusters are
discussed whereas the conclusions are given in the last section.

\section{Soft-core two- and three-body potentials}

As mentioned in the Introduction, we use the LM2M2 $^4$He-$^4$He potential
as the reference interaction, with the mass parameter fixed to 
$\hbar^2/m=43.281307~\text{(a.u.)}^2\,\text{K}$. In order to explore the 
Efimov-$(a^{-1},\kappa)$ plane, we have modified the LM2M2 interaction as following
\begin{equation}
  V_\lambda(r)=\lambda \cdot V_{\text{LM2M2}}(r)\,\, .
\label{mtbp}
\end{equation}
\begin{figure}
  \begin{center}
  \includegraphics[width=\linewidth]{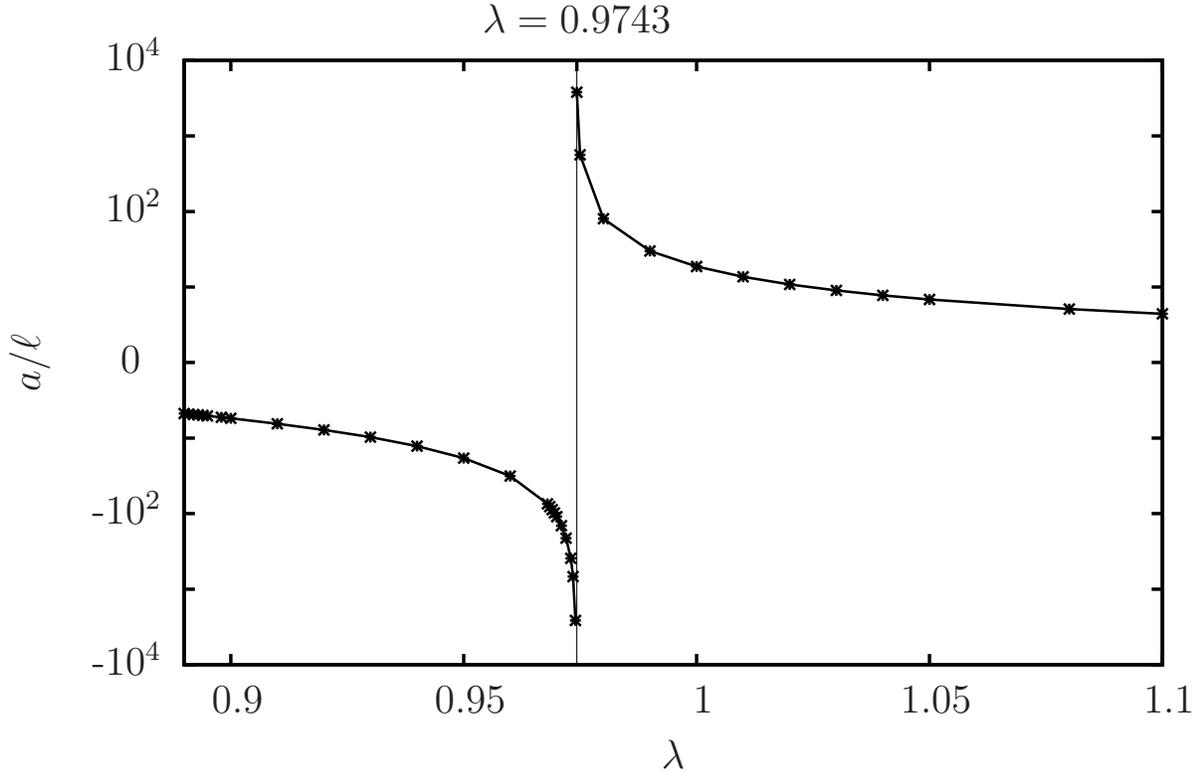}
  \end{center}
  \caption{The scattering length $a$, in units of $\ell$, as a function of the parameter $\lambda$,
    calculated with the modified LM2M2 potential $V_\lambda(r)$. The range of
    variation of $\lambda$ is between $\lambda=0.883$, which corresponds to the
    disappearance in the continuum of the excited three-body state, and
    $\lambda=1.1$. The unitary limit is obtained for $\lambda \approx 0.9743$.}
 \label{fig:plota}
\end{figure}
Examples of this strategy exist in the literature~\cite{esry:1996_phys.rev.a,barletta:2001_phys.rev.a}.
We have varied $\lambda$ from $\lambda=0.883$, where 
$a=a^0_-=-43.84$ a.u., up 
to $\lambda=1.1$ corresponding to $a=44.79$ a.u., as shown in
Fig~\ref{fig:plota}. The unitary limit 
is produced for $\lambda\approx 0.9743$. When $\lambda=1$ the values of the LM2M2
are recovered: $a=189.41$ a.u., $E_2$=-1.303 mK and $r_0=13.845$ a.u..

\begin{figure}
  \begin{center}
  \includegraphics[width=\linewidth]{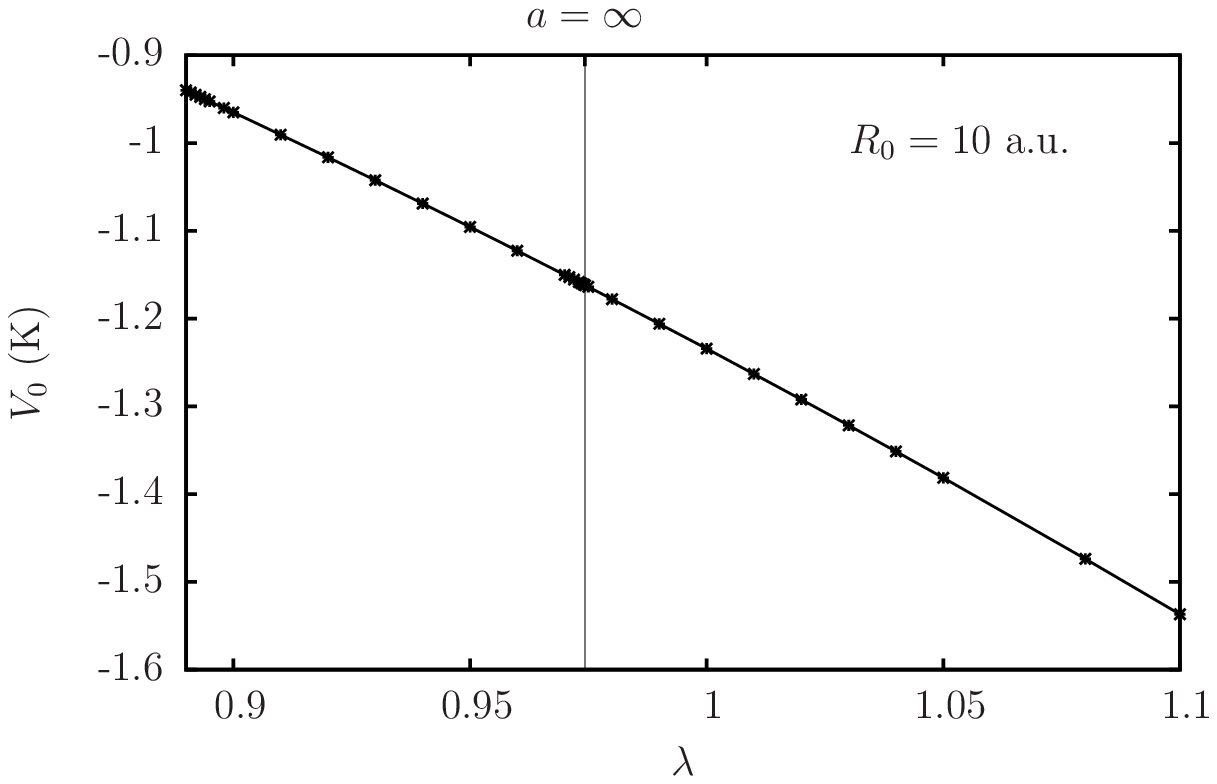}%
  \end{center}
  \caption{The strength $V_0$ of the gaussian two-body-potential as a 
function of the parameter $\lambda$.
     The values are tuned to reproduce the scattering length $a$ given
   by the modified LM2M2 potential $V_\lambda(r)$.}
 \label{fig:V0_vs_lambda}
\end{figure}

Following Refs.~\cite{nielsen:2001_phys.rep.,kievsky:2011_few-bodysyst.,gattobigio:2011_phys.rev.a} 
we have constructed an attractive two-body gaussian (TBG) potential
\begin{equation}
V(r)=V_0 \,\, {\rm e}^{-r^2/R_0^2}\,,
\label{eq:twobp}
\end{equation}
with range $R_0=10$~a.u., 
and we have varied the strength $V_0$ in order to reproduce the values of $a$ given
by $V_\lambda(r)$, as shown in Fig.~\ref{fig:V0_vs_lambda}. For $\lambda=1$ 
with the strength $V_0=-1.2343566$~K we reproduce the LM2M2 low-energy data,
$E_{2}=-1.303$ mK, $a=189.42$ a.u., and $r_0=13.80$ a.u.. 
\begin{figure}
  \begin{center}
  \includegraphics[width=\linewidth]{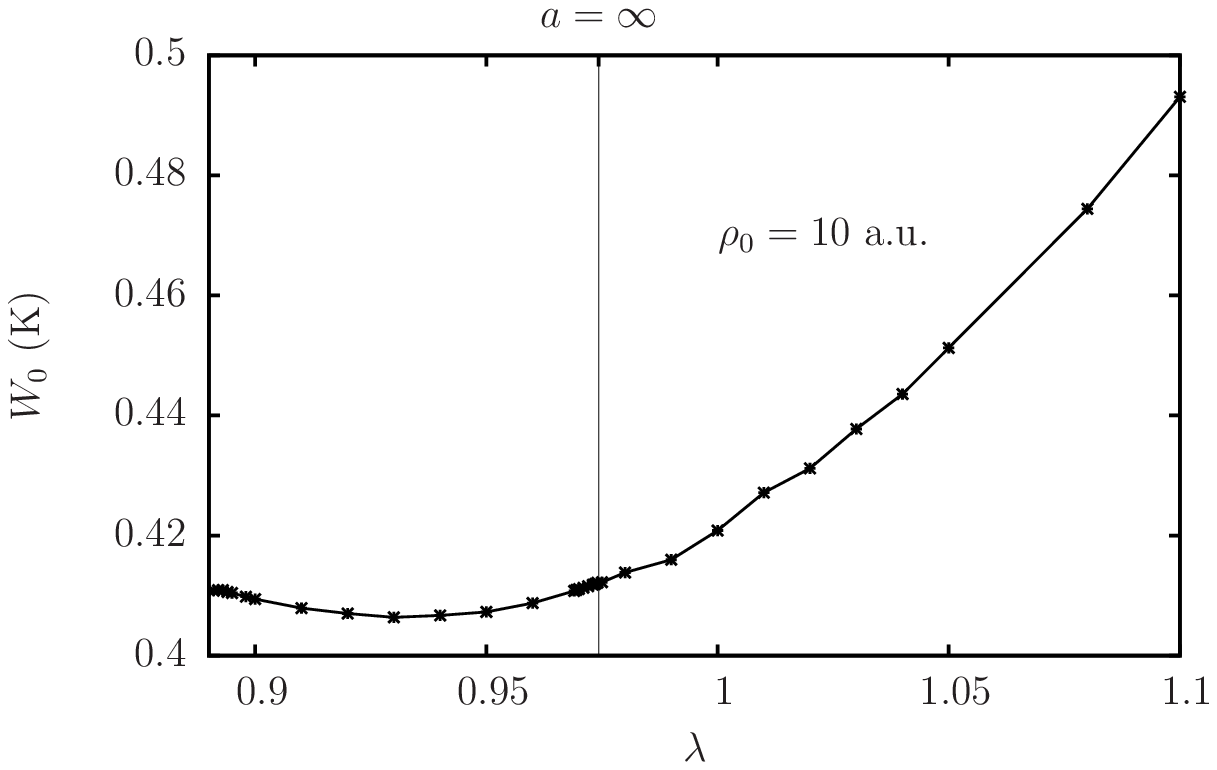}
  \end{center}
  \caption{The strength $W_0$  of the hypercentral three-body-potential
as a function of the parameter $\lambda$.
     The values are tuned to reproduce the three-body ground state $E_3^0$ given by
   the modified LM2M2 potential $V_\lambda(r)$.}
 \label{fig:W0_vs_lambda}
\end{figure}
The use of the TBG potential in the three-atom system produces a ground state binding
energy  appreciable deeper than the one calculated with $V_\lambda(r)$.
For example, at $\lambda=1$ the LM2M2 helium
trimer ground state binding energy is $126.4$ mK whereas the one obtained using 
the two-body-soft-core potential in  Eq.~(\ref{eq:twobp})
is $151.32$ mK. A smaller difference, though
still appreciable, can be observed in the first excited state. 

In order to have a closer description to the $A=3$ system obtained with the
modified LM2M2 potential, we introduce the following (repulsive)
hypercentral-three-body (H3B) interaction
\begin{equation}
W(\rho_{123})=W_0 \,\, {\rm e}^{-\rho^2_{123}/\rho^2_0}\,,
  \label{eq:hyptbf}
\end{equation}
with the strength $W_0$ tuned to reproduce the trimer energy $E_3^0$ obtained using
$V_\lambda(r)$ for all the explored values of $\lambda$, as shown in
Fig.~\ref{fig:W0_vs_lambda}.
Here $\rho^2_{123}=\frac{2}{3}(r^2_{12}+r^2_{23}+r^2_{31})$ is the hyperradius
of three particles 
and $\rho_0$ gives the range of the three-body force, or, in the spirit of 
EFT, the cut-off of the three-body-contact interaction; therefore, it is not
independent of $R_0$, which is the cut-off of the two-body-contact force, and 
in fact it should be $\rho_0=R_0$, as shown in
\cite{gattobigio:2011_phys.rev.a}, and so we fixed $\rho_0 = 10.0$~a.u.  
A different criterion to fix the three-body force was given in
Ref.~\cite{von_stecher:2010_j.phys.b:at.mol.opt.phys.} in which the condition
$\rho_0 \gg r_0$ has been used. In this case the (repulsive) three-body force is
used to push the trimer spectrum high in energy in order to verify as close as
possible the universal ratios $E^{n+1}_3/E^n_3={\rm e}^{-2\pi/s_0}$ already at
$n=0$.  With the LM2M2 interaction this relation is only approximate verified;
for $\lambda=1$ we have $E^0_3/E^1_3\approx 56$,  whereas at the unitary limit
$E^0_3/E^1_3\approx 525$, very close to the universal ratio.  

\begin{table}
\begin{center}
\begin{tabular}{@{}ccc}
\hline
\hline
potential & $E^{0}_{3}$ (mK) &  $E^{1}_{3}$ (mK)   \cr
\hline
\multicolumn{3}{c}{$\lambda=1$}          \cr
\cline{2-2}
 $ V_\lambda(r)$    &  -126.4  &  -2.27     \cr
 TBG &  -151.3  &  -2.48     \cr
 TBG+H3B   &  -126.4  &  -2.31     \cr
\hline
\multicolumn{3}{c}{$\lambda=0.9743$}   \cr
\cline{2-2}
 $ V_\lambda(r)$    &  -83.99  &  -0.16     \cr
 TBG &  -103.4  &  -0.20     \cr
 TBG+H3B   &  -83.99  &  -0.16     \cr
\hline
\hline
\end{tabular}
  \caption{The ground state $E^{0}_{3}$ and the excited state $E^{1}_{3}$
of the three-boson system calculated with the modified LM2M2 potential 
$V_\lambda(r)$, the TBG potential, and the TBG potential plus the H3B 
potential at $\lambda=1$, that corresponds to the original LM2M2 potential, and
in the unitary limit, $\lambda=0.9743$.}
\label{tab:table1}
\end{center}
\end{table}

\section{Three-body Efimov Plot}
The calculations for $A=3$ have been performed using the CHH expansion. Since
$V_\lambda(r)$ is obtained multiplying  the LM2M2 potential by a global
factor $\lambda$, it inherits the strong short range repulsion; in this case, a direct
use of the HH basis to compute the bound states is not feasible since it would
be necessary to include an enormous number of basis elements in the 
expansion~\cite{kievsky:1997_few-bodysyst}. The use of the CHH expansion circumvents this problem 
by the introduction of a correlation factor of the Jastrow type. The method is described in
Ref.~\cite{barletta:2001_phys.rev.a} and it allows to achieve similar
accuracy as other techniques.  As an example, in
Table~\ref{tab:table1} we show the results for the ground state $E_3^0$ and the
excited state $E_3^1$ at $\lambda=1$ (in this case the results of the
LM2M2 potential are recovered), and at the unitary limit ($\lambda=0.9743$).
These results have been obtained using the CHH basis up to
a value of the grand-angular momentum $K=160$. 

\begin{figure}
  \begin{center}
  \includegraphics[width=\linewidth]{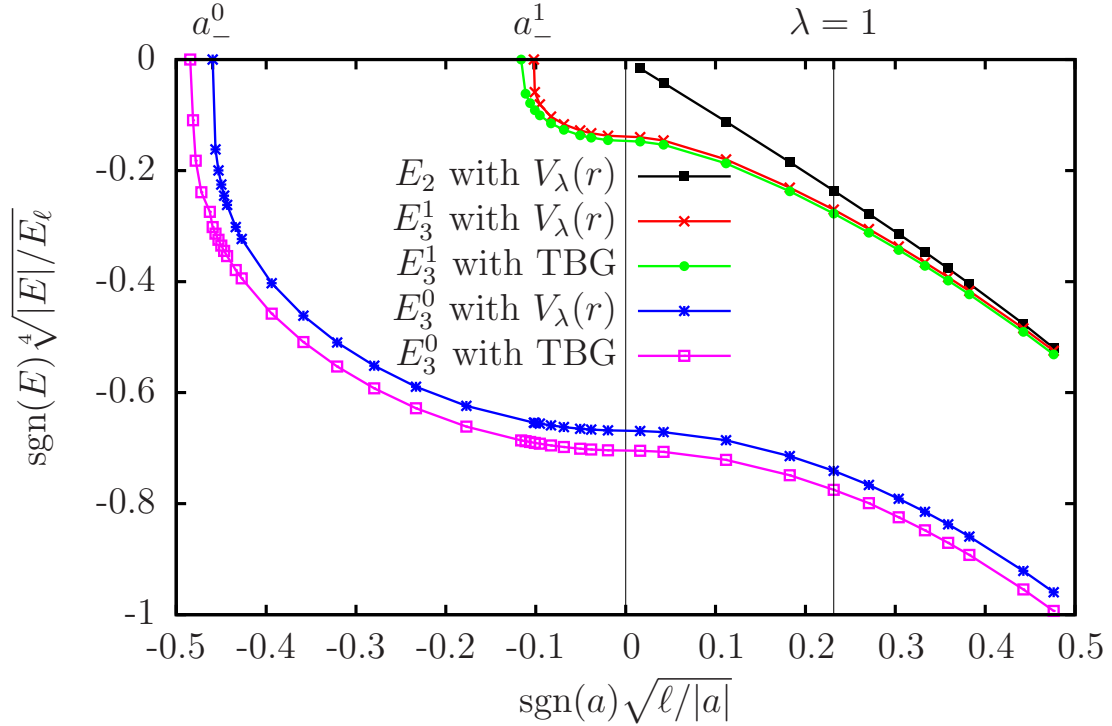}
 \end{center}
  \caption{(color online). Efimov plot for $A=3$. We report the  ground and excited 
 state energies, $E_3^0$ and $E_3^1$, in units of $E_\ell$, as a function of $\ell/a$,
 both for the modified LM2M2 
 potential  $V_\lambda(r)$, and
 for the TBG potential. Following the literature, we really draw the fourth root of
 the scaled energies as a function of the square root of the scaled-inverse
 scattering length; using this trick, the ratio between excited and ground
 energies is greatly reduced, allowing for the graphical representation of both
 curves on the same scale. We also report 
 the $A=2$ binding energy.}
 \label{fig:plot1}
\end{figure}

As a byproduct of the tuning procedure of the three-body strength $W_0$, we
have constructed, as was previously done for instance in
Refs.~\cite{esry:1996_phys.rev.a,naidon:2012_phys.rev.a}, 
the Efimov plot shown in Fig.~\ref{fig:plot1}. 
In the figure we report calculations of $E_3^0$ and $E_3^1$ as functions
of $a$ done both with the $V_\lambda(r)$ and the TBG potential. When the
TBG+H3B potential is used, the results coincide with those of $V_\lambda(r)$
and are not reported in the figure. In addition, we draw the dimer energy $E_2$, 
calculated using the $V_\lambda(r)$ potential. In order to show these quantities
together in the figure we have used the fourth root of the energy (in units of
$E_\ell$) as a function of the square root of $a^{-1}$ (in units of $\ell$).
In the region analyzed, the results are inside the Efimov window; in fact, the scattering 
length is still much larger than the natural length $\ell$, and the ground-state energy 
$E_3^0$ is above the natural value  $E_\ell$.

Looking in Fig.~\ref{fig:plot1} at negative values of $a$, 
it is possible to identify the value of the scattering
length $a_-^1$ at which the excited state $E_3^1$ disappears. 
For $V_\lambda(r)$, this value is $a_-^1\approx -975 \,$a.u.,
whereas using the TBG potential it results $a_-^1\approx -752  \,$a.u.. The
next interesting point appears at $a_-^0$, when the three-body cluster is no more
bound, so that $E_3^0$ approaches zero. Using $V_\lambda(r)$
this happens at $a_-^0\approx -48.1$ a.u., whereas using the
TBG potential alone it is $a_-^0\approx -43.3$ a.u..

The ratio $a_-^0/a_-^1$ has been predicted to have an universal value
${(a_-^0/a_-^1)}_{\text{theory}} = 22.7$~\cite{braaten:2006_physicsreports};  in
the only experiment which measures the two thresholds,
Ref.~\cite{pollack:2009_science}, the ratio is
${(a_-^0/a_-^1)}_{\text{experiment}} = 21.1$. In our case we obtain
${(a_-^0/a_-^1)}_{\text{TBG}} = 17.4$, and ${(a_-^0/a_-^1)}_{V_\lambda(r)} =
20.3$, which is closer to both theoretical and experimental values.  The
absolute position of $a_-^0$ is not predicted by the theory of Efimov physics and,
in that sense it can be considered as not an universal quantity; 
however, it has been the subject of
experimental measurements which give more or less the same value in units of
mean scattering length $\overline a = 0.955978 \, \ell/2$ for different atoms,
$a_-^0 = -(9\pm 1) \,\overline a$~\cite{berninger:2011_phys.rev.lett.} . In the
present calculations we obtain $({a_-^0})_{\text{TBG}} = - 8.9 \,\overline a$
and $({a_-^0})_{V_\lambda(r)} = - 9.9 \,\overline a $. 

\begin{figure}
  \begin{center}
  \includegraphics[width=\linewidth]{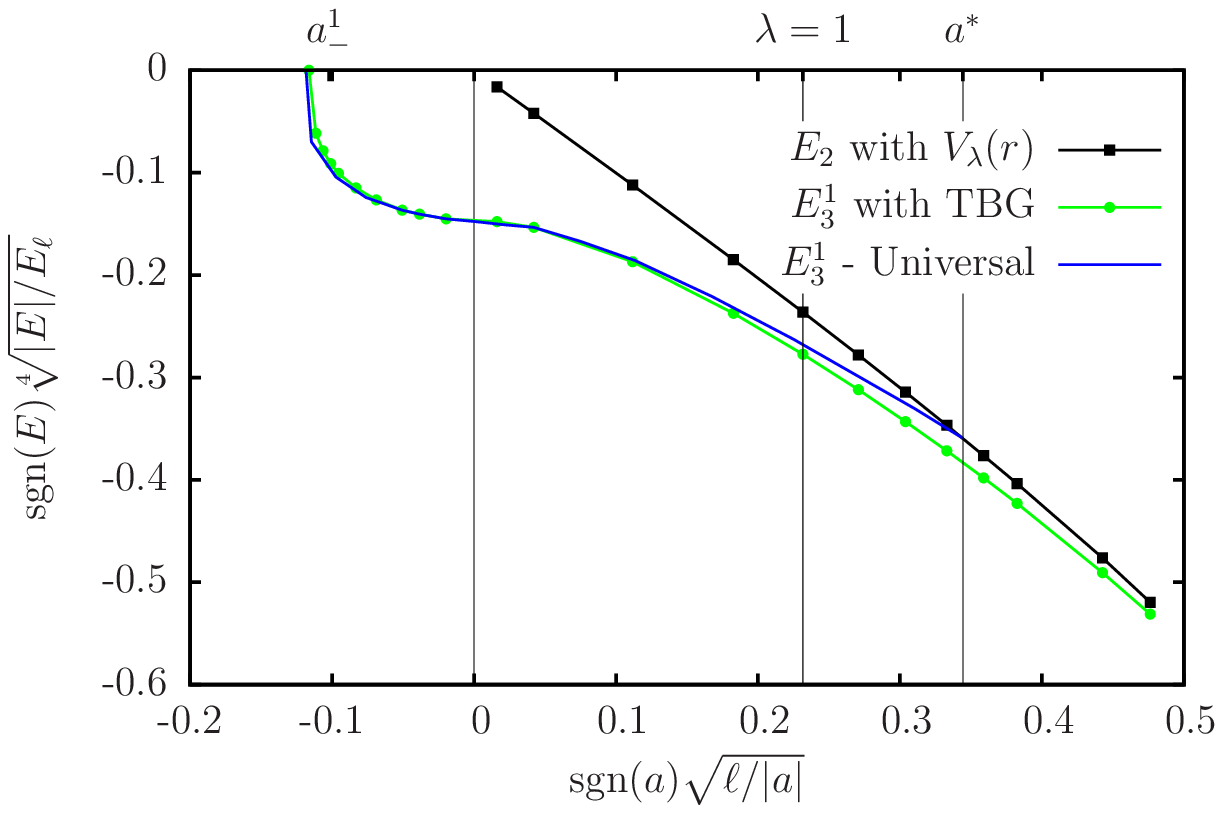}
  \includegraphics[width=\linewidth]{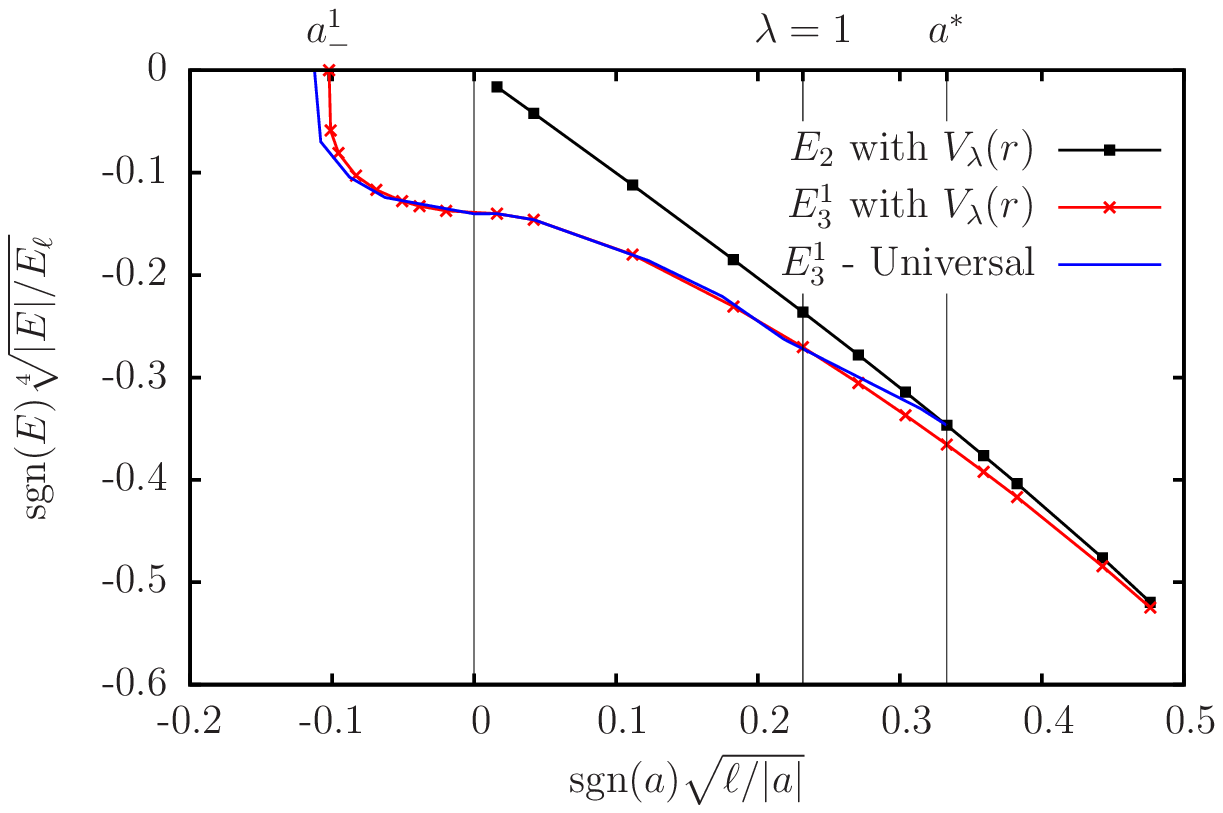}
  \end{center}
  \caption{(color online). Comparison between the
  excited three-body energy $E_3^1$ and the
theoretical universal value given by Eq.~(\ref{eq:uni3}),
using the TBG potential
(upper panel), and the modified LM2M2 potential $V_\lambda(r)$ (lower panel). 
The two-body energy $E_2$ calculated with $V_\lambda(r)$ is also shown. 
The calculated and theoretical curves agree around the resonant
limit, and the differences close to the values  $a^*$ and $a^1_-$ are due to 
effective-range corrections. The most remarkable difference is represented by
the fact that the calculated $E_3^1$ does not cross the atom-dimer threshold.}
\label{fig:plotnew} 
\end{figure}

In addition, we discuss the universal character of the shallow state $E^1_3$.
Using the Efimov's radial law~\cite{efimov:1971_sov.j.nucl.phys.} it is possible
to obtain an equation for this trimer binding energy as a function of $a$. It
reads

\begin{equation}
 E_3^1+\frac{\hbar^2}{ma^2}=\exp{[\Delta(\xi)/s_0]}\frac{\hbar^2\kappa^2_*}{m}\,,
\label{eq:uni3}
\end{equation}
where $\kappa_*$ is the wave number corresponding to the energy
$E^1_3=\hbar^2\kappa_*^2/m  = 0.156$~mK
at the
resonant limit and $\tan{\xi}=-(m E_3^1/\hbar^2)^{1/2}a$. The function
$\Delta(\xi)$ is universal and a parametrization in the range $[-\pi,-\pi/4]$ is
given in Ref.~\cite{braaten:2006_physicsreports}. It verifies $\Delta(-\pi/2)=0$
and, from the very precise result $\Delta(-\pi/4)=6.02730678199$ and
$\Delta(-\pi)\approx -0.89$~\cite{braaten:2006_physicsreports}, it is possible to
determine the values $a^*$ (at which $E_3^1=E_2$) and $a^1_-$ (at which
$E_3^1=0$). In order to analyze the universal character of the calculated
energies $E_3^1$ using the TBG and TBG+H3B potential models, in
Fig.~\ref{fig:plotnew} we compare them to the values of Eq.(\ref{eq:uni3}). By
construction the energies $E_3^1$ at the resonant limit coincide with those of
Eq.(\ref{eq:uni3}).  It is possible to see that the calculated energies using
the TBG potential, Fig.~\ref{fig:plotnew} upper panel, and TBG+H3B potential,
Fig.~\ref{fig:plotnew} lower panel, reproduce the universal behaviour close to
the resonant limit. The small differences observed at finite values of $a$, 
especially  close to the
critical values $a^*$ and $a^1_-$, are due to effective-range corrections which
are automatically included in our approach. Moreover, $E_3^1$ does not
disappear in the atom-dimer continuum at $a^*$ but follows very close the $E_2$
curve from below. 

\begin{figure}
  \begin{center}
    \includegraphics[width=\linewidth]{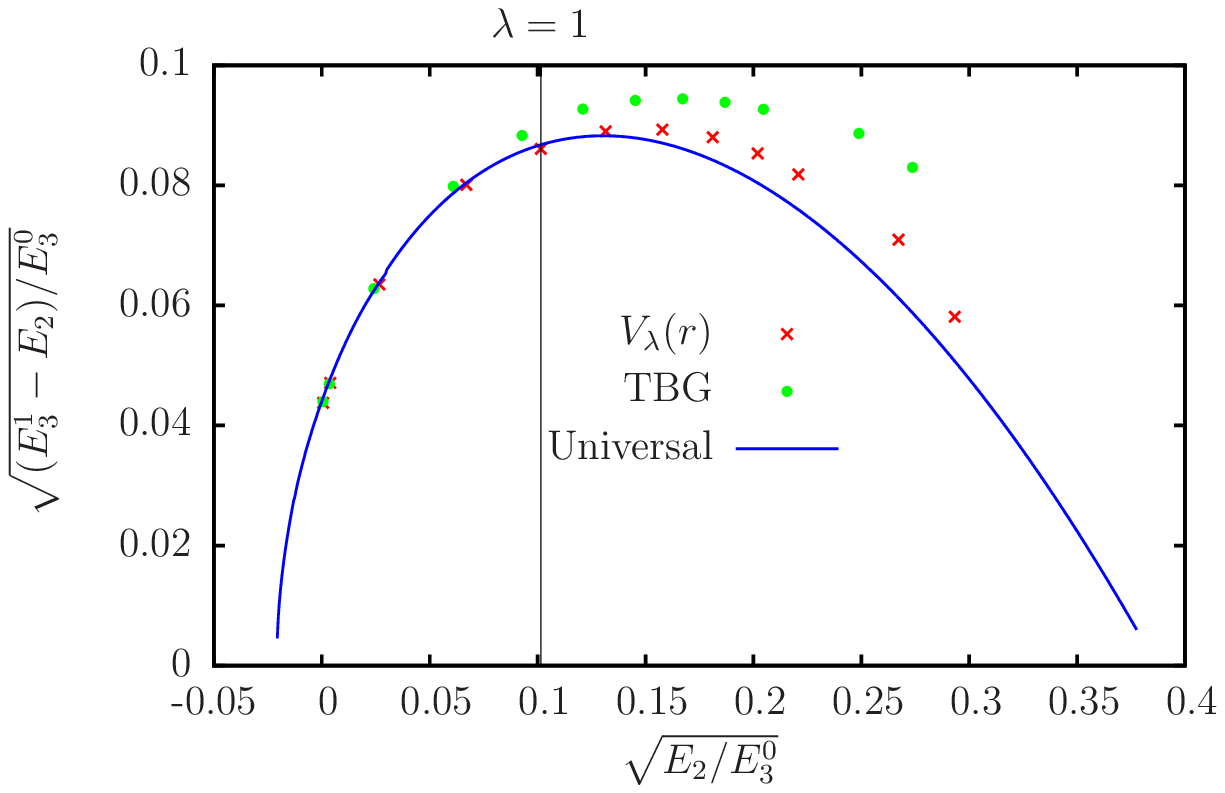}
  \end{center}
  \caption{(color online). Correlations between the ground $E_3^0$ and excited
    $E_3^1$ states of the trimer. We compare the universal correlation, 
    obtained by means of Eq.~(\ref{eq:uni3}), to the calculations made
    using the full $V_\lambda(r)$ potential and the TBG potential.  
  The agreement is good close to the unitary point, where the dimer energy $E_2$
  is small. The deviations become significant when the finite effective-range
effects become non-negligible.}
  \label{fig:universality}
\end{figure}

To conclude, we further analyze the universality looking at the correlations between 
the three-body ground and excited states, as has been proposed in
Ref.~\cite{frederico:1999_phys.rev.a}. In Fig.~\ref{fig:universality} we trace
the square root of the excited-trimer energy, measured from the two-body
dimer, in units of the trimer-ground state energy, as a function of the dimer energy,
always in units of trimer-ground state energy. The Efimov's universal-radial
law Eq.~(\ref{eq:uni3}) gives the universal curve in this plot; we see that as
far as the dimer is very shallow, the calculated points are very close to the 
universal curve. They depart from it when corrections due  both to finite scattering
length and to non-zero effective range become sizeable. This non-universal effect is more 
important for the TBG case, probably due to the lack of the three-body
corrections.

\section{Efimov plot for $A=4,5,6$ clusters}

The calculations for the $A>3$ systems are performed using
NSHH basis. The method has been recently used to describe up to
six nucleons interacting through a central
potential~\cite{gattobigio:2011_phys.rev.c,gattobigio:2009_few-bodysyst.,gattobigio:2011_j.phys.:conf.ser.}
and six bosons using a two-body
plus a three-body force~\cite{gattobigio:2011_phys.rev.a}. 
The Hamiltonian matrix is obtained using the following orthonormal basis
\begin{equation}
  \langle\rho\,\Omega\,|\,m\,[K]\rangle =
  \bigg(\beta^{(\alpha+1)/2}\sqrt{\frac{m!}{(\alpha+m)!}}\,
  L^{(\alpha)}_m(\beta\rho)
  \,{\text e}^{-\beta\rho/2}\bigg)
  {\cal Y}^{LM}_{[K]}(\Omega_N)  \,,
  \label{mhbasis}
\end{equation}
where $L^{(\alpha)}_m(\beta\rho)$ is a Laguerre polynomial with $\alpha=3N-1$
($N=A-1$) and $\beta$ a variational non-linear parameter. The function ${\cal
Y}^{LM}_{[K]}(\Omega_N)$ are the HH functions with grand-angular momentum $K$,
and total angular momenta $L$ and magnetic number $M$.
The Hamiltonian matrix is not constructed, but using properties of HH is
expressed as an algebraic combination of sparse matrices, allowing for an efficient
research of the lowest eigenvectors/eigenvalues. A full discussion of the NSHH
method is given in
Refs.~\cite{gattobigio:2011_phys.rev.c,gattobigio:2011_phys.rev.a}.

\begin{figure}
  \begin{center}
  \includegraphics[width=\linewidth]{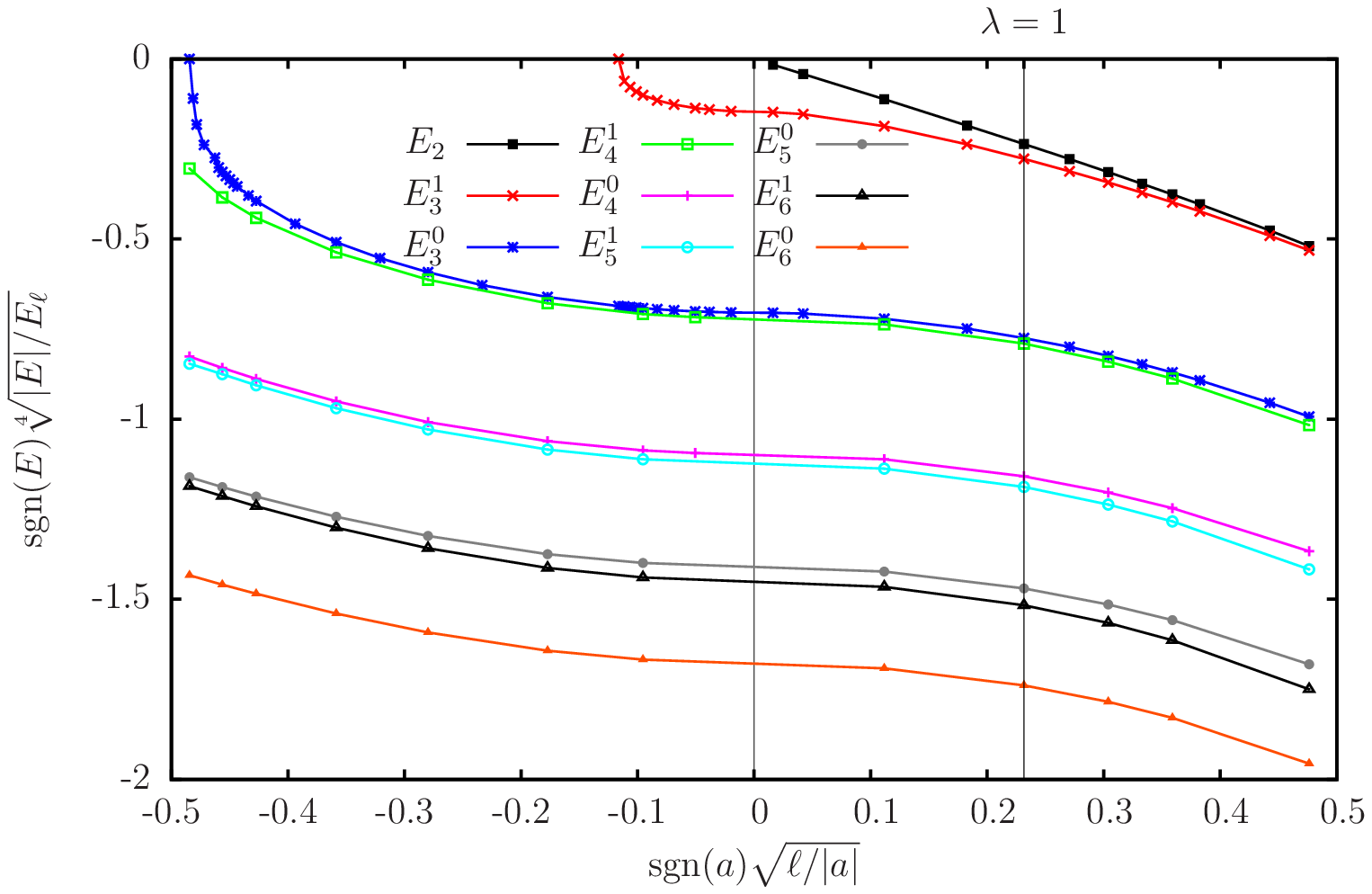}
  \includegraphics[width=\linewidth]{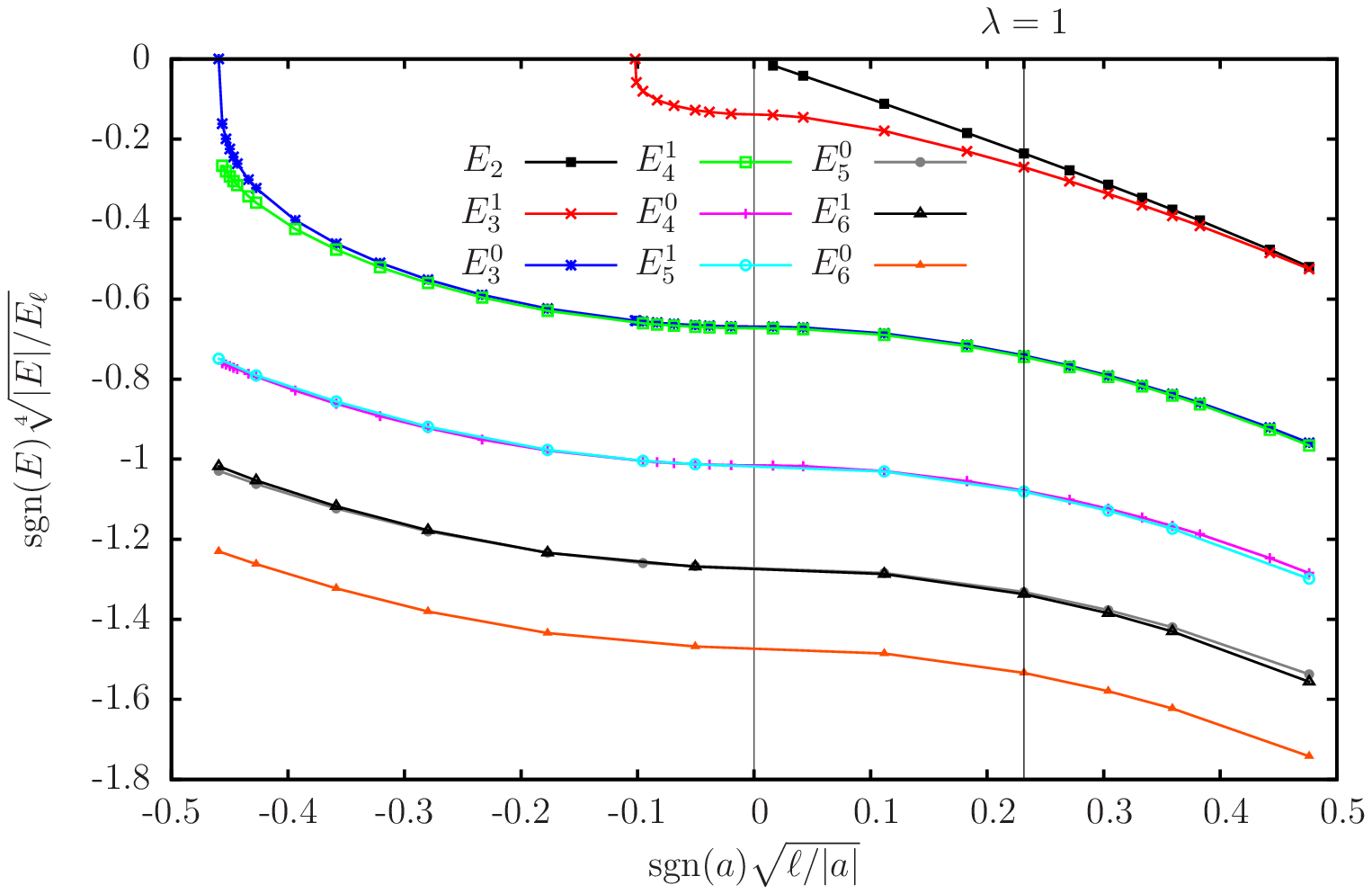}
  \end{center}
  \caption{(color online). Energies of the $A=3-6$ ground and excited 
 states, $E_A^0,E_A^1$, as a function of $a^{-1}$, using
 the two-body gaussian potential (upper panel), and using the two-body plus the 
 hypercentral three-body force (lower panel). In both panels we also give 
 the two-body ground-state energy $E_2$ calculated with the LM2M2 potential.}
 \label{fig:plot2}
\end{figure}

\begin{figure}
  \begin{center}
  \includegraphics[width=\linewidth]{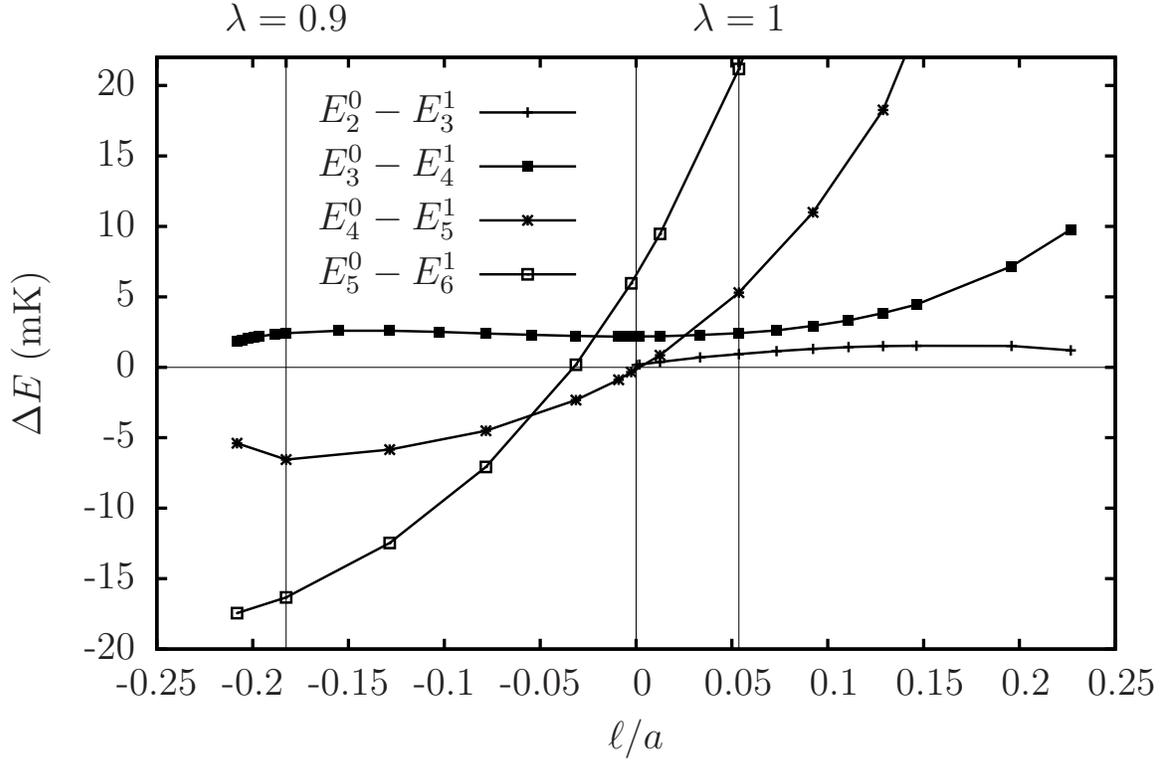}
  \end{center}
  \caption{The difference $\Delta E=E^0_A-E^1_{A+1}$ for the indicated cases as
 a function of the inverse of $a$ for the TBG+H3B potential. 
 The particular cases at $\lambda=1$ and $0.9$
 as well as at the unitary limit are indicated as vertical lines.}
 \label{fig:efimov_differences}
\end{figure}

After solving the $A=3$ problem for bound states, used to fix the strength of
the H3B force, we have diagonalized the Hamiltonian for $A=4,5,6$ bodies using
the TBG and TBG+H3B potentials. The results are given in Fig.~\ref{fig:plot2} in
two scaled-$(a^{-1},\kappa)$ plots, one obtained with the two-body potential
alone (upper panel) and one with the two-body plus three-body interactions
(lower panel). 
In the first case, with only the TBG potential, we observe that the spectrum of the
systems $A=4,5,6$ presents two bound states, one deep and one shallow, for all
values of $a$ studied. When the repulsive three-body force is included,
the spectrum moves up and we can observe
that the excited state $E_A^1$ disappears for $A=5,6$ 
for negative values of the scattering length
as $a$ approaches $a^1_-$. 
This fact is better shown in Fig.~\ref{fig:efimov_differences}, where the
differences $E^0_A-E^1_{A+1}$ have been plotted as a functions of $\ell/a$. Whereas
the differences $E^0_2-E^1_3$ and $E^0_3-E^1_4$ are positive along the
whole range, indicating that the states $E^1_3$ and $E^1_4$ 
are bound, the differences $E^0_4-E^1_5$ and $E^0_5-E^1_6$ result
negatives as $a$ goes to the negative region, so at some value of $a$ the 
excited states $E^1_5$ and $E^1_6$ are no more bound. The determination
of the point where the transition happens can be determined by looking at the
convergence of the states $E^1_5$ and $E^1_6$, as can be seen in
Table~\ref{tab:convergence} where we report the convergence pattern
using the TBG+H3B potential for $A=4,5,6$ at the point $\lambda = 0.9$ where both 
$E^1_5$ and $E^1_6$ are not bound. They remain above the $E^0_4$ and $E^0_5$ threshold
respectively. In the table we have shown the three maximum values 
of $K$ considered in the present calculations. 

\begin{figure}
  \begin{center}
  \includegraphics[width=\linewidth]{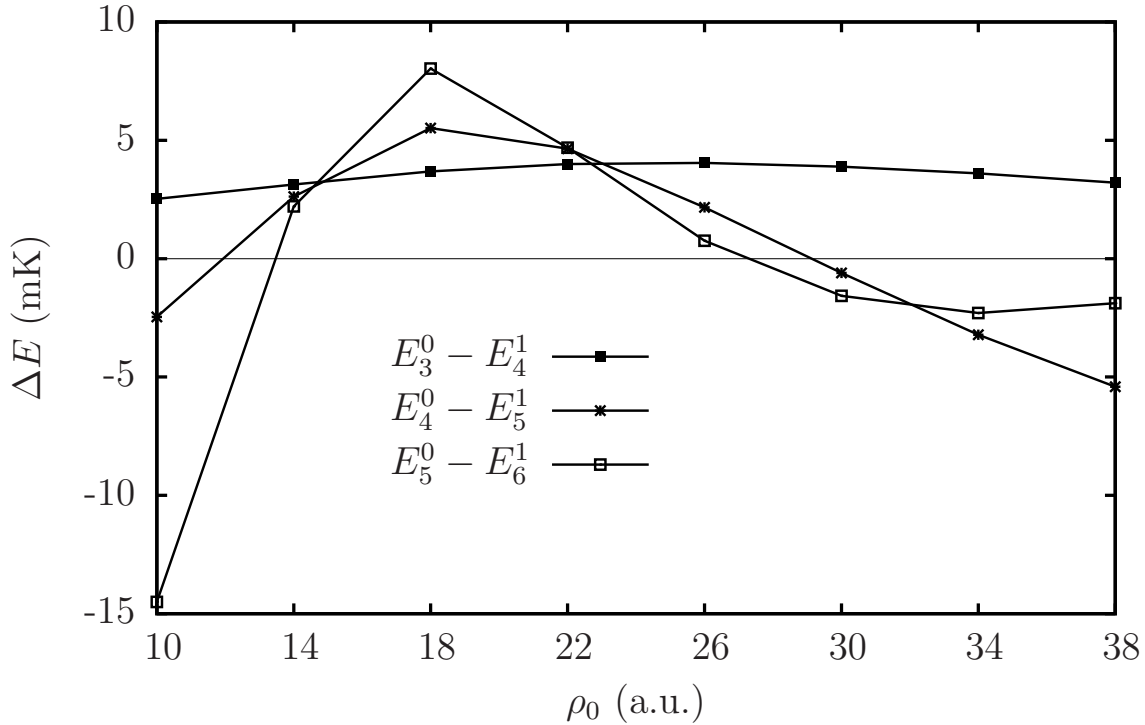}
  \end{center}
  \caption{The difference $\Delta E=E^0_A-E^1_{A+1}$ at $\lambda=0.9$ as
 a function of $\rho_0$ for the TBG+H3B potential.}
 \label{fig:rho0}
\end{figure}

Moreover, the fact that the states  $E^1_5$ and $E^1_6$ are bound or not
depends also on the range of the three-body force $\rho_0$.
In order to analyze this relation, we have varied $\rho_0$  at
$\lambda=0.9$ as well as at the unitary limit. For each value of $\rho_0$ the
strength of the three-body potential has been fixed to reproduce the trimer
binding energy $E_3^0$ as before. The results for $A=4,5,6$ at $\lambda=0.9$ are
shown in Fig.~\ref{fig:rho0}. As can be seen, the excited states are recovered
as bound states for values of $\rho_0\approx 18\, {\rm a.u.}$. 

\begin{figure}
  \begin{center}
  \includegraphics[width=\linewidth]{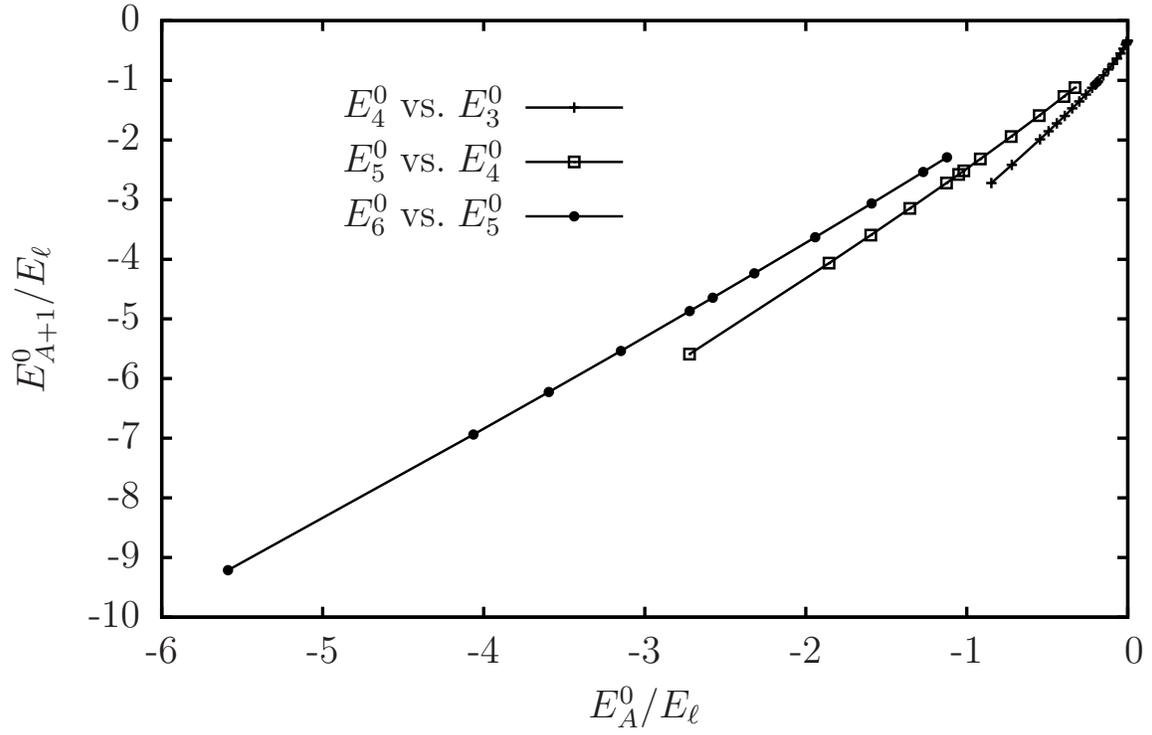}
  \includegraphics[width=\linewidth]{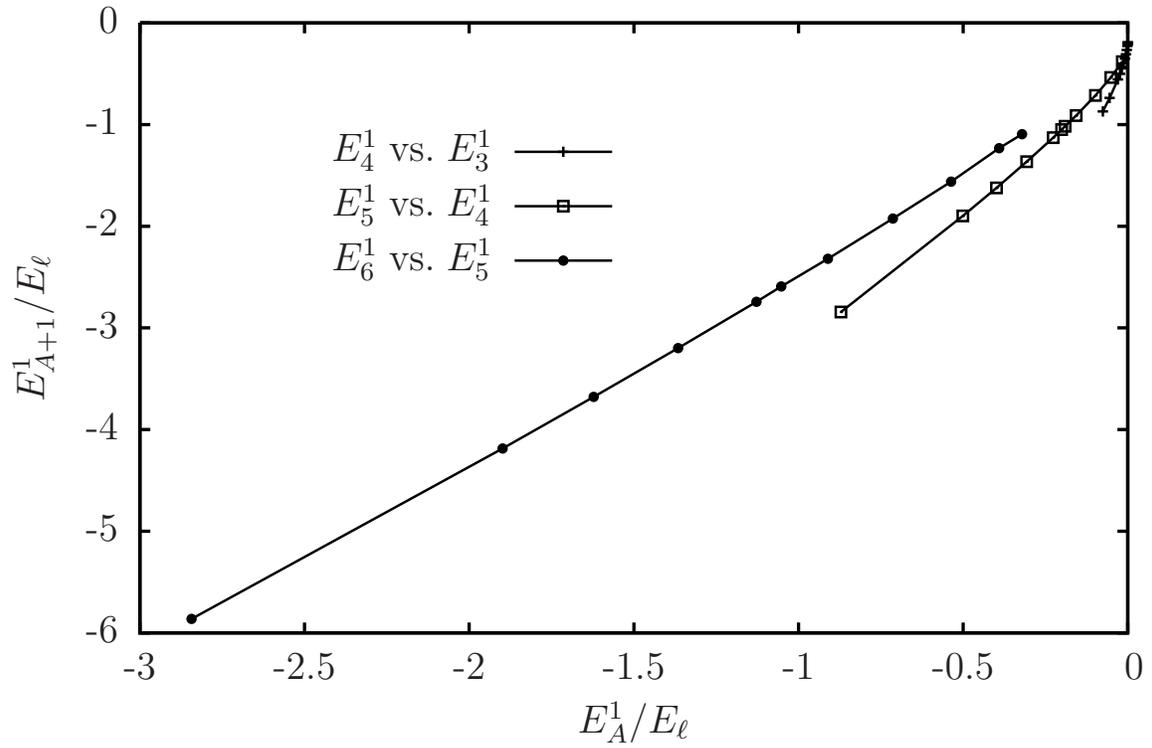}
  \end{center}
  \caption{Relation between $E^0_{A+1}$ and $E^0_A$ (upper panel), and between
  $E^1_{A+1}$ and $E^1_A$ (lower panel) for the indicated cases
 obtained with the TBG+H3B potential along the $(a^{-1},\kappa)$ plane.}
 \label{fig:tjon}
\end{figure}

To make contact with the analysis of Ref.~\cite{hadizadeh:2011_phys.rev.lett.} we have calculated
$\sqrt{|E_4^1-E_3^0|/|E_4^0|}=0.070$ and $\sqrt{|E_3^0|/|E_4^0|}=0.434$ at the unitary limit.
These two values correspond to a point in the plot given in Fig.1 of that reference
lying very close to line giving the relation of these two quantities at the
unitary limit. 
In addition, in Fig.~\ref{fig:tjon} we analyze the relation between
$E^0_{A+1}$ (upper panel) and $E^1_{A+1}$ (lower panel) with $E^0_A$ 
and $E^1_A$ respectively, as a function of the scattering length for $A=4,5,6$.  
We can
observe a linear dependence in all cases except for a small curvature in the
$E^0_4$ $vs.$ $E^0_3$ and $E^1_4$ $vs.$ $E^1_3$ curves close to the point in 
which $E^0_3$ goes to zero.
These curves display the universal character of these clusters as
their spectrum is determined by two parameters, $a$ and $E^0_3$. 

Besides, the universal ratios $E^0_{A}/E^0_3$ and $E^1_{A+1}/E^0_A$ can be
studied at $\lambda=1$.  They are: $E^0_4/E^0_3=4.5$, $E^0_5/E^0_3=10.4$,
$E^0_6/E^0_3=18.4$ and $E^1_4/E^0_3=1.020$, $E^1_5/E^0_4=1.009$,
$E^1_6/E^0_5=1.016$.  These ratios are in close agreement to those obtained in
the literature~\cite{von_stecher:2010_j.phys.b:at.mol.opt.phys.}. At the unitary
limit the ratios $E^0_{A}/E^0_3$ move a little bit from the universal values
showing some dependence on the form of the soft potential whereas the ratios
$E^1_{A+1}/E^0_A$ show stability. At $\lambda=0.9743$ they are:
$E^0_4/E^0_3=5.3$, $E^0_5/E^0_3=13.0$, $E^0_6/E^0_3=23.4$ and
$E^1_4/E^0_3=1.026$, $E^1_5/E^0_4=1.004$, $E^1_6/E^0_5=1.006$.

\begin{figure}
  \begin{center}
  \includegraphics[width=\linewidth]{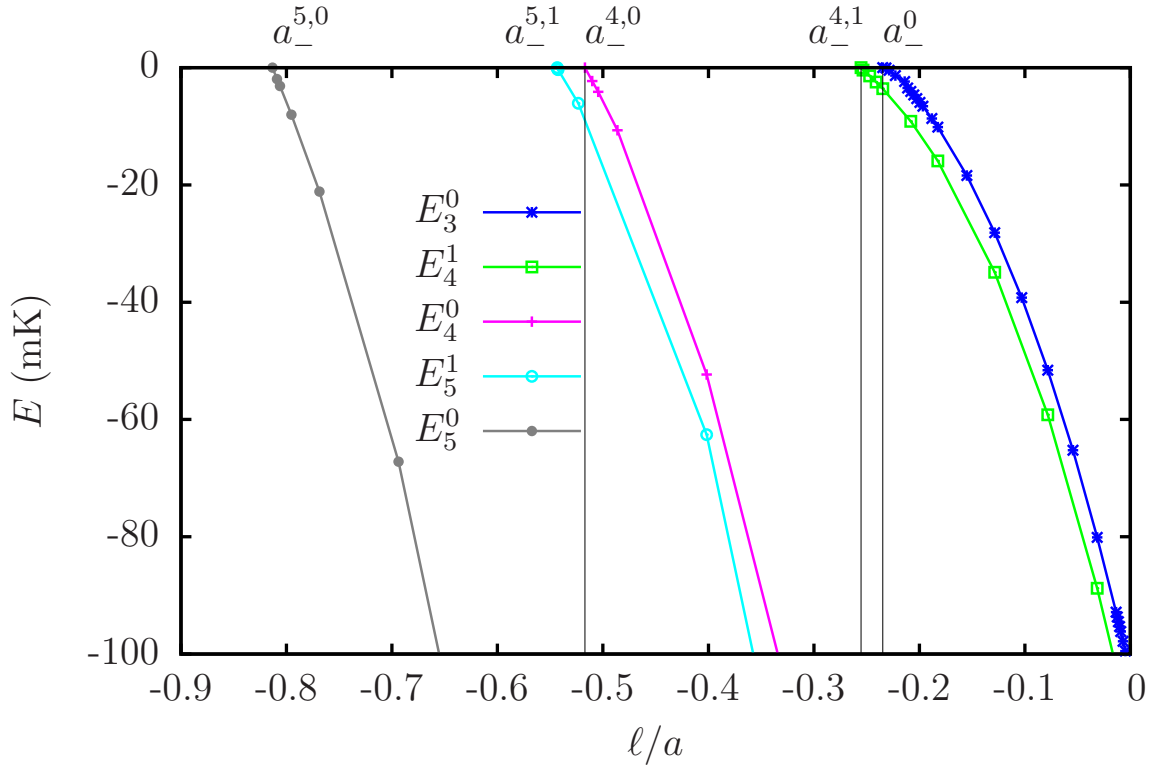}
  \end{center}
  \caption{(color online). Energies of the states $E_3^0$, $E_4^1$, $E_4^0$,
    $E_5^1$, and $E_5^0$ as a function of $a^{-1}$ for negative values of the
  scattering length close to the continuum threshold obtained using the TBG
  potential. The four-particle thresholds are $a_-^{4,0}\approx -19.6$~a.u., and
   $a_-^{4,1}\approx -39.8$~a.u. The five-particle thresholds are 
   $a_-^{5,0}\approx -12.5$~a.u., and
   $a_-^{5,1}\approx -18.7$~a.u.}
 \label{fig:thresholds}
\end{figure}

Finally, using the TBG potential, we have extended the calculations for $A=4$
and $A=5$ systems up to the four- and five-particle thresholds in order to
calculate the ratios between the different thresholds and to compare our results
to previous calculations and experimental outcomes.  Our results are summarized
in Fig.~\ref{fig:thresholds}.  Denoting with $a_-^{4,0}$ and  $a_-^{4,1}$ the
four-particle thresholds of the ground and excited state respectively, we have
$a_-^{4,1}\approx -39.8$~a.u., and $a_-^{4,0}\approx -19.6$~a.u.; equivalently,
with respect to the three-particle threshold $a_-^0$ we get
$(a_-^{4,1}/a_-^0)_{\text{TBG}}\approx 0.92$ for the excited state, and
$(a_-^{4,0}/a_-^0)_{\text{TBG}}\approx 0.45$ for the ground states. These values
agree very well with what measured in the
experiments~\cite{pollack:2009_science,zaccanti:2009_natphys,ferlaino:2009_phys.rev.lett.}
and with what predicted by the
theory~\cite{von_stecher:2009_natphys,von_stecher:2010_j.phys.b:at.mol.opt.phys.,deltuva:2012_}
The five-particle thresholds  read 
 $a_-^{5,0}\approx -12.5$~a.u. for the ground state and  $a_-^{5,1}\approx
-18.7$~a.u. for the excited state. Equivalently, the ratios with respect the four-particle
threshold  are  $(a_-^{5,0}/a_-^{4,0})_{\text{TBG}}\approx 0.64$ and
$(a_-^{5,1}/a_-^{4,0})_{\text{TBG}}\approx 0.95$, still in agreement with
previous theoretical
prediction~\cite{von_stecher:2010_j.phys.b:at.mol.opt.phys.} and with recent 
experiments~\cite{zenesini:2012_}.

\begin{table}
  \caption{Convergence of the binding energies as a function of the grand angular 
quantum number $K$ using the TBG+H3B potential at $\lambda = 0.9$,
   $R_0=10$~a.u., and $\rho_0=10$~a.u., for clusters of $A$ particles. We also report the 
   number $N_\text{HH}$ of hyperspherical basis elements
 corresponding to a given $K$.}
  \label{tab:convergence}
\begin{center}
  \begin{tabular*}{0.9\linewidth}{@{\extracolsep{\fill}}c c c c c}
  \hline 
  \hline 
  A \rule{0pt}{12pt} &  $K$ & $N_{\text{HH}}$ & $E^{0}_A$ (mK) & $E^{1}_A$(mK) \\
  \hline 
  \multirow{3}{*}{4} 
          & 36   & 33649  & -166.25945 & -6.55041\\
          & 38   & 42504  & -166.25949 & -6.79163\\
          & 40   & 53130  & -166.25951 & -6.99574\\
  \hline                  
  \multirow{3}{*}{5}   
              & 26    & 448800   & -532.75811 & -161.96737 \\
              & 28    & 724812   & -532.75828 & -162.98374 \\
              & 30    & 1139544  & -532.75834 & -163.79689 \\
  \hline                  
  \multirow{3}{*}{6}
        & 18  & 709410    & -1063.8276 & -513.50956\\
        & 20  & 1628328   & -1063.8311 & -516.42712\\
        & 22  & 3527160   & -1063.8322 & -518.25341\\
   \hline
   \hline
\end{tabular*}
\end{center}
\end{table}

\section{Conclusions}

In the present paper we have discussed the spectrum of bosonic
systems up to six particles interacting through a two-body potential
having a large two-body scattering length (with respect to
the effective range). The three-body scale has been fixed using a scaled
Helium-Helium potential. The scope was to extend previous studies on the
Efimov physics done in the three- and four-body systems. We have
observed that, similarly to the four-body system, the five- and
six-body systems present two bound states, one deep and one
shallow. It seems that this type of spectrum is to some extend
universal depending only on the condition $a \gg r_0$ in the two-body
system. This condition produces a geometrical series of bound states
in the three-body system and, attached to each of these states, a
two-level spectrum has been showed to appear in the four-body
system~\cite{platter:2004_phys.rev.a,von_stecher:2009_natphys,deltuva:2011_few-bodysyst.}. 
However they are true bound
states only in correspondence to the lowest trimer level. The other
states appear as resonances embedded in the continuum of four particles.
It is possible to introduce a repulsive three-body
force that eliminate all the trimer states below one specific level.
In this way, the two level spectrum of the four-body system attached to
this trimer ground state will become true bound states. Also the universal
character of the spectrum will be more evident as the repulsive three-body
force will push more and more the particles faraway.

Though the analysis of the bosonic spectrum can follow the strategy
illustrated above, in the present paper we follow a different one
based on the physics introduced by the two-body potential. In nuclear
systems as well as in many atomic systems the two-body interaction
has a sharp repulsion at short range followed by a very weak attractive
part that produces very shallow dimers as for example the deuteron or
the two-helium molecule. The particles are located in the asymptotic
region and do not feel the details of the interaction. Therefore, we
can introduce a soft potential to be considered as a 
regularized-two-body contact term in an EFT 
approximation of the original potential~\cite{lepage:1997_}. 
In the three-body sector, a three-body-contact term is required to
reproduce the ground-state-binding energy of three-particles
introduced here by means of a gaussian-hypercentral three-body force.
Using this potential model we have calculated the three-body spectrum
and we have analyzed its universal character comparing the energy
of the shallow state $E^1_3$ to the Efimov's equation for the binding
energy. Furthermore we have identified the critical values $a^0_-$ and
$a^1_-$ at which $E^0_3=0$ and $E^1_3=0$ respectively.
Successively we have calculated the four-, five-, and six-body spectra and we
have observed the two-level structure. The EFT approach is better
adapted to describe shallow states and this is confirmed by the
close result obtained for $E_4^1=-128.8$ mK at $\lambda=1$ and
the LM2M2 result $E_4^1=-127.9$ mK, very recently published~\cite{hiyama:2012_phys.rev.a}.

The universal character of the structure of these clusters has been studied
using Tjon lines, that means  the relation between $E^0_{A+1}$ and $E^0_A$, and between
$E^1_{A+1}$ and $E^1_A$. As illustrated
in Fig.~\ref{fig:tjon}, we have obtained an almost linear relation between
$E^0_{A+1}$ and $E^0_A$ (upper panel) and between $E^1_{A+1}$ and $E^1_A$ (lower
panel) in the region from $\lambda=1$ to $0.9$. As the energy
of the cluster, $E^0_A$ or $E^1_A$, tends to zero the linear relation is lost. 

Since we are
describing the lowest bound states, some universal ratios are only approximately
verified, though not very far from the values quoted by other groups in $A=3,4$.
However, in the simple case of TBG we have extended our calculations for $A=4$
and $A=5$ up to the four- and five-particle continuum threshold in order to
calculate the ratios between the thresholds: the values we obtain in the
four-body case, $(a_-^{4,1}/a_-^0)_{\text{TBG}}\approx 0.92$ and 
$(a_-^{4,0}/a_-^0)_{\text{TBG}}\approx 0.45$, and in the five-body system
$(a_-^{5,0}/a_-^{4,0})_{\text{TBG}}\approx 0.64$ and 
$(a_-^{5,1}/a_-^{4,0})_{\text{TBG}}\approx 0.95$, are in accord with the ratios
that have been previously predicted~\cite{von_stecher:2011_phys.rev.lett.} 
and measured~\cite{zenesini:2012_} in literature.

Another interesting aspect is the uncertainty introduced by the cutoff in the
hypercentral three-body force. We have observed that with the most natural
choice $\rho_0=R$ the shallow states $E_5^1$ and $E_6^1$ result unbound in the
last part of the curves. They cross the respective threshold $E_4^0$ and
$E_5^0$. Increasing $\rho_0$ they result bound again around $\rho_0\approx 18\,
{a.u.}$. Increasing further $\rho_0$ they become again unbound. This last
analysis is somehow inconclusive as to really  understand the cutoff dependence
we need to vary both cutoff $R_0$ and $\rho_0$ in a coherent way; the dependence
on the cutoff will eventually reflect the leading order nature of the potential
we are using,  pointing to the necessity of going to a higher order in the EFT
expansion~\cite{lepage:1997_}.  Studies along this line are at present under
consideration as well as the analysis of the two-level spectrum for cluster with
more than six particles.

\newpage

\newpage


\begin{thebibliography}{36}%
\makeatletter
\providecommand \@ifxundefined [1]{%
 \@ifx{#1\undefined}
}%
\providecommand \@ifnum [1]{%
 \ifnum #1\expandafter \@firstoftwo
 \else \expandafter \@secondoftwo
 \fi
}%
\providecommand \@ifx [1]{%
 \ifx #1\expandafter \@firstoftwo
 \else \expandafter \@secondoftwo
 \fi
}%
\providecommand \natexlab [1]{#1}%
\providecommand \enquote  [1]{``#1''}%
\providecommand \bibnamefont  [1]{#1}%
\providecommand \bibfnamefont [1]{#1}%
\providecommand \citenamefont [1]{#1}%
\providecommand \href@noop [0]{\@secondoftwo}%
\providecommand \href [0]{\begingroup \@sanitize@url \@href}%
\providecommand \@href[1]{\@@startlink{#1}\@@href}%
\providecommand \@@href[1]{\endgroup#1\@@endlink}%
\providecommand \@sanitize@url [0]{\catcode `\\12\catcode `\$12\catcode
  `\&12\catcode `\#12\catcode `\^12\catcode `\_12\catcode `\%12\relax}%
\providecommand \@@startlink[1]{}%
\providecommand \@@endlink[0]{}%
\providecommand \url  [0]{\begingroup\@sanitize@url \@url }%
\providecommand \@url [1]{\endgroup\@href {#1}{\urlprefix }}%
\providecommand \urlprefix  [0]{URL }%
\providecommand \Eprint [0]{\href }%
\providecommand \doibase [0]{http://dx.doi.org/}%
\providecommand \selectlanguage [0]{\@gobble}%
\providecommand \bibinfo  [0]{\@secondoftwo}%
\providecommand \bibfield  [0]{\@secondoftwo}%
\providecommand \translation [1]{[#1]}%
\providecommand \BibitemOpen [0]{}%
\providecommand \bibitemStop [0]{}%
\providecommand \bibitemNoStop [0]{.\EOS\space}%
\providecommand \EOS [0]{\spacefactor3000\relax}%
\providecommand \BibitemShut  [1]{\csname bibitem#1\endcsname}%
\let\auto@bib@innerbib\@empty
\bibitem [{\citenamefont {Braaten}\ and\ \citenamefont
  {Hammer}(2006)}]{braaten:2006_physicsreports}%
  \BibitemOpen
  \bibfield  {author} {\bibinfo {author} {\bibfnamefont {Eric}\ \bibnamefont
  {Braaten}}\ and\ \bibinfo {author} {\bibfnamefont {{H.-W.}}\ \bibnamefont
  {Hammer}},\ }\bibfield  {title} {\enquote {\bibinfo {title} {Universality in
  few-body systems with large scattering length},}\ }\href {\doibase
  10.1016/j.physrep.2006.03.001} {\bibfield  {journal} {\bibinfo  {journal}
  {Physics Reports}\ }\textbf {\bibinfo {volume} {428}},\ \bibinfo {pages}
  {259--390} (\bibinfo {year} {2006})}\BibitemShut {NoStop}%
\bibitem [{\citenamefont {Greene}(2010)}]{greene:2010_phys.today}%
  \BibitemOpen
  \bibfield  {author} {\bibinfo {author} {\bibfnamefont {Chris~H.}\
  \bibnamefont {Greene}},\ }\bibfield  {title} {\enquote {\bibinfo {title}
  {Universal insights from few-body land},}\ }\href {\doibase
  10.1063/1.3366239} {\bibfield  {journal} {\bibinfo  {journal} {Phys. Today}\
  }\textbf {\bibinfo {volume} {63}},\ \bibinfo {pages} {40} (\bibinfo {year}
  {2010})}\BibitemShut {NoStop}%
\bibitem [{\citenamefont {Ferlaino}\ and\ \citenamefont
  {Grimm}(2010)}]{ferlaino:2010_physics}%
  \BibitemOpen
  \bibfield  {author} {\bibinfo {author} {\bibfnamefont {Francesca}\
  \bibnamefont {Ferlaino}}\ and\ \bibinfo {author} {\bibfnamefont {Rudolf}\
  \bibnamefont {Grimm}},\ }\bibfield  {title} {\enquote {\bibinfo {title}
  {Forty years of efimov physics: How a bizarre prediction turned into a hot
  topic},}\ }\href {\doibase 10.1103/Physics.3.9} {\bibfield  {journal}
  {\bibinfo  {journal} {Physics}\ }\textbf {\bibinfo {volume} {3}},\ \bibinfo
  {pages} {9} (\bibinfo {year} {2010})}\BibitemShut {NoStop}%
\bibitem [{\citenamefont {Efimov}(1970)}]{efimov:1970_phys.lett.b}%
  \BibitemOpen
  \bibfield  {author} {\bibinfo {author} {\bibfnamefont {V}~\bibnamefont
  {Efimov}},\ }\bibfield  {title} {\enquote {\bibinfo {title} {Energy levels
  arising from resonant two-body forces in a three-body system},}\ }\href
  {\doibase 10.1016/0370-2693(70)90349-7} {\bibfield  {journal} {\bibinfo
  {journal} {Phys. Lett. B}\ }\textbf {\bibinfo {volume} {33}},\ \bibinfo
  {pages} {563--564} (\bibinfo {year} {1970})}\BibitemShut {NoStop}%
\bibitem [{\citenamefont {Efimov}(1971)}]{efimov:1971_sov.j.nucl.phys.}%
  \BibitemOpen
  \bibfield  {author} {\bibinfo {author} {\bibfnamefont {V}~\bibnamefont
  {Efimov}},\ }\bibfield  {title} {\enquote {\bibinfo {title} {Weak bound
  states of three resonantly interacting particles},}\ }\href@noop {}
  {\bibfield  {journal} {\bibinfo  {journal} {Sov. J. Nucl. Phys.}\ }\textbf
  {\bibinfo {volume} {12}},\ \bibinfo {pages} {589} (\bibinfo {year} {1971})},\
  \bibinfo {note} {{[Yad.} Fiz. 12, 1080–1090 (1970)].}\BibitemShut {Stop}%
\bibitem [{\citenamefont {Platter}\ \emph {et~al.}(2004)\citenamefont
  {Platter}, \citenamefont {Hammer},\ and\ \citenamefont
  {Meißner}}]{platter:2004_phys.rev.a}%
  \BibitemOpen
  \bibfield  {author} {\bibinfo {author} {\bibfnamefont {L.}~\bibnamefont
  {Platter}}, \bibinfo {author} {\bibfnamefont {{H.-W.}}\ \bibnamefont
  {Hammer}}, \ and\ \bibinfo {author} {\bibfnamefont {{Ulf-G.}}\ \bibnamefont
  {Meißner}},\ }\bibfield  {title} {\enquote {\bibinfo {title} {Four-boson
  system with short-range interactions},}\ }\href {\doibase
  10.1103/PhysRevA.70.052101} {\bibfield  {journal} {\bibinfo  {journal} {Phys.
  Rev. A}\ }\textbf {\bibinfo {volume} {70}},\ \bibinfo {pages} {052101}
  (\bibinfo {year} {2004})}\BibitemShut {NoStop}%
\bibitem [{\citenamefont {von Stecher}\ \emph {et~al.}(2009)\citenamefont {von
  Stecher}, \citenamefont {{D’Incao}},\ and\ \citenamefont
  {Greene}}]{von_stecher:2009_natphys}%
  \BibitemOpen
  \bibfield  {author} {\bibinfo {author} {\bibfnamefont {J.}~\bibnamefont {von
  Stecher}}, \bibinfo {author} {\bibfnamefont {J.~P.}\ \bibnamefont
  {{D’Incao}}}, \ and\ \bibinfo {author} {\bibfnamefont {Chris~H.}\
  \bibnamefont {Greene}},\ }\bibfield  {title} {\enquote {\bibinfo {title}
  {Signatures of universal four-body phenomena and their relation to the efimov
  effect},}\ }\href {\doibase 10.1038/nphys1253} {\bibfield  {journal}
  {\bibinfo  {journal} {Nat Phys}\ }\textbf {\bibinfo {volume} {5}},\ \bibinfo
  {pages} {417--421} (\bibinfo {year} {2009})}\BibitemShut {NoStop}%
\bibitem [{\citenamefont {Deltuva}\ \emph {et~al.}(2011)\citenamefont
  {Deltuva}, \citenamefont {Lazauskas},\ and\ \citenamefont
  {Platter}}]{deltuva:2011_few-bodysyst.}%
  \BibitemOpen
  \bibfield  {author} {\bibinfo {author} {\bibfnamefont {A.}~\bibnamefont
  {Deltuva}}, \bibinfo {author} {\bibfnamefont {R.}~\bibnamefont {Lazauskas}},
  \ and\ \bibinfo {author} {\bibfnamefont {L.}~\bibnamefont {Platter}},\
  }\bibfield  {title} {\enquote {\bibinfo {title} {Universality in {Four-Body}
  scattering},}\ }\href {\doibase 10.1007/s00601-011-0227-8} {\bibfield
  {journal} {\bibinfo  {journal} {{Few-Body} Syst.}\ }\textbf {\bibinfo
  {volume} {51}},\ \bibinfo {pages} {235--247} (\bibinfo {year}
  {2011})}\BibitemShut {NoStop}%
\bibitem [{\citenamefont {Hammer}\ and\ \citenamefont
  {Platter}(2007)}]{hammer:2007_eur.phys.j.a}%
  \BibitemOpen
  \bibfield  {author} {\bibinfo {author} {\bibfnamefont {H.~{-W.}}\
  \bibnamefont {Hammer}}\ and\ \bibinfo {author} {\bibfnamefont
  {L.}~\bibnamefont {Platter}},\ }\bibfield  {title} {\enquote {\bibinfo
  {title} {Universal properties of the four-body system with large scattering
  length},}\ }\href {\doibase 10.1140/epja/i2006-10301-8} {\bibfield  {journal}
  {\bibinfo  {journal} {Eur. Phys. J. A}\ }\textbf {\bibinfo {volume} {32}},\
  \bibinfo {pages} {113--120} (\bibinfo {year} {2007})}\BibitemShut {NoStop}%
\bibitem [{\citenamefont {Kolganova}\ \emph {et~al.}(1998)\citenamefont
  {Kolganova}, \citenamefont {Motovilov},\ and\ \citenamefont
  {Sofianos}}]{kolganova:1998_j.phys.b}%
  \BibitemOpen
  \bibfield  {author} {\bibinfo {author} {\bibfnamefont {E~A}\ \bibnamefont
  {Kolganova}}, \bibinfo {author} {\bibfnamefont {A~K}\ \bibnamefont
  {Motovilov}}, \ and\ \bibinfo {author} {\bibfnamefont {S~A}\ \bibnamefont
  {Sofianos}},\ }\bibfield  {title} {\enquote {\bibinfo {title} {Three-body
  configuration space calculations with hard-core potentials},}\ }\href
  {http://stacks.iop.org/0953-4075/31/i=6/a=014} {\bibfield  {journal}
  {\bibinfo  {journal} {J. Phys. B}\ }\textbf {\bibinfo {volume} {31}},\
  \bibinfo {pages} {1279} (\bibinfo {year} {1998})}\BibitemShut {NoStop}%
\bibitem [{\citenamefont {Nielsen}\ \emph {et~al.}(2001)\citenamefont
  {Nielsen}, \citenamefont {Fedorov}, \citenamefont {Jensen},\ and\
  \citenamefont {Garrido}}]{nielsen:2001_phys.rep.}%
  \BibitemOpen
  \bibfield  {author} {\bibinfo {author} {\bibfnamefont {E.}~\bibnamefont
  {Nielsen}}, \bibinfo {author} {\bibfnamefont {D.~V.}\ \bibnamefont
  {Fedorov}}, \bibinfo {author} {\bibfnamefont {A.~S.}\ \bibnamefont {Jensen}},
  \ and\ \bibinfo {author} {\bibfnamefont {E.}~\bibnamefont {Garrido}},\
  }\bibfield  {title} {\enquote {\bibinfo {title} {The three-body problem with
  short-range interactions},}\ }\href {\doibase DOI:
  10.1016/S0370-1573(00)00107-1} {\bibfield  {journal} {\bibinfo  {journal}
  {Phys. Rep.}\ }\textbf {\bibinfo {volume} {347}},\ \bibinfo {pages} {373 --
  459} (\bibinfo {year} {2001})}\BibitemShut {NoStop}%
\bibitem [{\citenamefont {Barletta}\ and\ \citenamefont
  {Kievsky}(2001)}]{barletta:2001_phys.rev.a}%
  \BibitemOpen
  \bibfield  {author} {\bibinfo {author} {\bibfnamefont {P.}~\bibnamefont
  {Barletta}}\ and\ \bibinfo {author} {\bibfnamefont {A.}~\bibnamefont
  {Kievsky}},\ }\bibfield  {title} {\enquote {\bibinfo {title} {Variational
  description of the helium trimer using correlated hyperspherical harmonic
  basis functions},}\ }\href {\doibase 10.1103/PhysRevA.64.042514} {\bibfield
  {journal} {\bibinfo  {journal} {Phys. Rev. A}\ }\textbf {\bibinfo {volume}
  {64}},\ \bibinfo {pages} {042514} (\bibinfo {year} {2001})}\BibitemShut
  {NoStop}%
\bibitem [{\citenamefont {Lewerenz}(1997)}]{lewerenz:1997_j.chem.phys.}%
  \BibitemOpen
  \bibfield  {author} {\bibinfo {author} {\bibfnamefont {Marius}\ \bibnamefont
  {Lewerenz}},\ }\bibfield  {title} {\enquote {\bibinfo {title} {Structure and
  energetics of small helium clusters: Quantum simulations using a recent
  perturbational pair potential},}\ }\href {\doibase 10.1063/1.473501}
  {\bibfield  {journal} {\bibinfo  {journal} {J. Chem. Phys.}\ }\textbf
  {\bibinfo {volume} {106}},\ \bibinfo {pages} {4596} (\bibinfo {year}
  {1997})}\BibitemShut {NoStop}%
\bibitem [{\citenamefont {Hiyama}\ and\ \citenamefont
  {Kamimura}(2012)}]{hiyama:2012_phys.rev.a}%
  \BibitemOpen
  \bibfield  {author} {\bibinfo {author} {\bibfnamefont {E.}~\bibnamefont
  {Hiyama}}\ and\ \bibinfo {author} {\bibfnamefont {M.}~\bibnamefont
  {Kamimura}},\ }\bibfield  {title} {\enquote {\bibinfo {title} {Variational
  calculation of {{\textasciicircum}{4}He} tetramer ground and excited states
  using a realistic pair potential},}\ }\href {\doibase
  10.1103/PhysRevA.85.022502} {\bibfield  {journal} {\bibinfo  {journal} {Phys.
  Rev. A}\ }\textbf {\bibinfo {volume} {85}},\ \bibinfo {pages} {022502}
  (\bibinfo {year} {2012})}\BibitemShut {NoStop}%
\bibitem [{\citenamefont {Timofeyuk}(2008)}]{timofeyuk:2008_phys.rev.c}%
  \BibitemOpen
  \bibfield  {author} {\bibinfo {author} {\bibfnamefont {N.}~\bibnamefont
  {Timofeyuk}},\ }\bibfield  {title} {\enquote {\bibinfo {title} {Improved
  procedure to construct a hyperspherical basis for the n-body problem:
  Application to bosonic systems},}\ }\href {\doibase
  10.1103/PhysRevC.78.054314} {\bibfield  {journal} {\bibinfo  {journal} {Phys.
  Rev. C}\ }\textbf {\bibinfo {volume} {78}},\ \bibinfo {pages} {054314}
  (\bibinfo {year} {2008})}\BibitemShut {NoStop}%
\bibitem [{\citenamefont {Kievsky}\ \emph {et~al.}(2011)\citenamefont
  {Kievsky}, \citenamefont {Garrido}, \citenamefont {{Romero-Redondo}},\ and\
  \citenamefont {Barletta}}]{kievsky:2011_few-bodysyst.}%
  \BibitemOpen
  \bibfield  {author} {\bibinfo {author} {\bibfnamefont {A.}~\bibnamefont
  {Kievsky}}, \bibinfo {author} {\bibfnamefont {E.}~\bibnamefont {Garrido}},
  \bibinfo {author} {\bibfnamefont {C.}~\bibnamefont {{Romero-Redondo}}}, \
  and\ \bibinfo {author} {\bibfnamefont {P.}~\bibnamefont {Barletta}},\
  }\bibfield  {title} {\enquote {\bibinfo {title} {The helium trimer with
  {Soft-Core} potentials},}\ }\href {\doibase 10.1007/s00601-011-0226-9}
  {\bibfield  {journal} {\bibinfo  {journal} {{Few-Body} Syst.}\ }\textbf
  {\bibinfo {volume} {51}},\ \bibinfo {pages} {259--269} (\bibinfo {year}
  {2011})}\BibitemShut {NoStop}%
\bibitem [{\citenamefont {Gattobigio}\ \emph
  {et~al.}(2011{\natexlab{a}})\citenamefont {Gattobigio}, \citenamefont
  {Kievsky},\ and\ \citenamefont {Viviani}}]{gattobigio:2011_phys.rev.a}%
  \BibitemOpen
  \bibfield  {author} {\bibinfo {author} {\bibfnamefont {M.}~\bibnamefont
  {Gattobigio}}, \bibinfo {author} {\bibfnamefont {A.}~\bibnamefont {Kievsky}},
  \ and\ \bibinfo {author} {\bibfnamefont {M.}~\bibnamefont {Viviani}},\
  }\bibfield  {title} {\enquote {\bibinfo {title} {Spectra of helium clusters
  with up to six atoms using soft-core potentials},}\ }\href {\doibase
  10.1103/PhysRevA.84.052503} {\bibfield  {journal} {\bibinfo  {journal} {Phys.
  Rev. A}\ }\textbf {\bibinfo {volume} {84}},\ \bibinfo {pages} {052503}
  (\bibinfo {year} {2011}{\natexlab{a}})}\BibitemShut {NoStop}%
\bibitem [{\citenamefont {Aziz}\ and\ \citenamefont
  {Slaman}(1991)}]{aziz:1991_j.chem.phys.}%
  \BibitemOpen
  \bibfield  {author} {\bibinfo {author} {\bibfnamefont {Ronald~A.}\
  \bibnamefont {Aziz}}\ and\ \bibinfo {author} {\bibfnamefont {Martin~J.}\
  \bibnamefont {Slaman}},\ }\bibfield  {title} {\enquote {\bibinfo {title} {An
  examination of ab initio results for the helium potential energy curve},}\
  }\href {\doibase 10.1063/1.460139} {\bibfield  {journal} {\bibinfo  {journal}
  {J. Chem. Phys.}\ }\textbf {\bibinfo {volume} {94}},\ \bibinfo {pages} {8047}
  (\bibinfo {year} {1991})}\BibitemShut {NoStop}%
\bibitem [{\citenamefont {Lepage}(1997)}]{lepage:1997_}%
  \BibitemOpen
  \bibfield  {author} {\bibinfo {author} {\bibfnamefont {Peter}\ \bibnamefont
  {Lepage}},\ }in\ \href@noop {} {\emph {\bibinfo {booktitle} {Nuclear Physics:
  Proceedings of the {VIII} Jorge André Swieca Summer School, 1995}}},\
  \bibinfo {editor} {edited by\ \bibinfo {editor} {\bibnamefont {{C. A.
  Bertulani, et al.}}}}\ (\bibinfo  {publisher} {World Scientific, Singapore,
  1997},\ \bibinfo {year} {1997})\ p.\ \bibinfo {pages} {135},\ \bibinfo {note}
  {{arXiv:nucl-th/9706029}}\BibitemShut {NoStop}%
\bibitem [{\citenamefont {Gattobigio}\ \emph
  {et~al.}(2009{\natexlab{a}})\citenamefont {Gattobigio}, \citenamefont
  {Kievsky}, \citenamefont {Viviani},\ and\ \citenamefont
  {Barletta}}]{gattobigio:2009_phys.rev.a}%
  \BibitemOpen
  \bibfield  {author} {\bibinfo {author} {\bibfnamefont {M.}~\bibnamefont
  {Gattobigio}}, \bibinfo {author} {\bibfnamefont {A.}~\bibnamefont {Kievsky}},
  \bibinfo {author} {\bibfnamefont {M.}~\bibnamefont {Viviani}}, \ and\
  \bibinfo {author} {\bibfnamefont {P.}~\bibnamefont {Barletta}},\ }\bibfield
  {title} {\enquote {\bibinfo {title} {Harmonic hyperspherical basis for
  identical particles without permutational symmetry},}\ }\href {\doibase
  10.1103/PhysRevA.79.032513} {\bibfield  {journal} {\bibinfo  {journal} {Phys.
  Rev. A}\ }\textbf {\bibinfo {volume} {79}},\ \bibinfo {pages} {032513}
  (\bibinfo {year} {2009}{\natexlab{a}})}\BibitemShut {NoStop}%
\bibitem [{\citenamefont {Gattobigio}\ \emph
  {et~al.}(2009{\natexlab{b}})\citenamefont {Gattobigio}, \citenamefont
  {Kievsky}, \citenamefont {Viviani},\ and\ \citenamefont
  {Barletta}}]{gattobigio:2009_few-bodysyst.}%
  \BibitemOpen
  \bibfield  {author} {\bibinfo {author} {\bibfnamefont {M.}~\bibnamefont
  {Gattobigio}}, \bibinfo {author} {\bibfnamefont {A.}~\bibnamefont {Kievsky}},
  \bibinfo {author} {\bibfnamefont {M.}~\bibnamefont {Viviani}}, \ and\
  \bibinfo {author} {\bibfnamefont {P.}~\bibnamefont {Barletta}},\ }\bibfield
  {title} {\enquote {\bibinfo {title} {Non-symmetrized basis function for
  identical particles},}\ }\href {\doibase 10.1007/s00601-009-0045-4}
  {\bibfield  {journal} {\bibinfo  {journal} {{Few-Body} Syst.}\ }\textbf
  {\bibinfo {volume} {45}},\ \bibinfo {pages} {127--131} (\bibinfo {year}
  {2009}{\natexlab{b}})}\BibitemShut {NoStop}%
\bibitem [{\citenamefont {Gattobigio}\ \emph
  {et~al.}(2011{\natexlab{b}})\citenamefont {Gattobigio}, \citenamefont
  {Kievsky},\ and\ \citenamefont {Viviani}}]{gattobigio:2011_phys.rev.c}%
  \BibitemOpen
  \bibfield  {author} {\bibinfo {author} {\bibfnamefont {M.}~\bibnamefont
  {Gattobigio}}, \bibinfo {author} {\bibfnamefont {A.}~\bibnamefont {Kievsky}},
  \ and\ \bibinfo {author} {\bibfnamefont {M.}~\bibnamefont {Viviani}},\
  }\bibfield  {title} {\enquote {\bibinfo {title} {Nonsymmetrized
  hyperspherical harmonic basis for an a-body system},}\ }\href {\doibase
  10.1103/PhysRevC.83.024001} {\bibfield  {journal} {\bibinfo  {journal} {Phys.
  Rev. C}\ }\textbf {\bibinfo {volume} {83}},\ \bibinfo {pages} {024001}
  (\bibinfo {year} {2011}{\natexlab{b}})}\BibitemShut {NoStop}%
\bibitem [{\citenamefont {von
  Stecher}(2010)}]{von_stecher:2010_j.phys.b:at.mol.opt.phys.}%
  \BibitemOpen
  \bibfield  {author} {\bibinfo {author} {\bibfnamefont {Javier}\ \bibnamefont
  {von Stecher}},\ }\bibfield  {title} {\enquote {\bibinfo {title} {Weakly
  bound cluster states of efimov character},}\ }\href {\doibase
  10.1088/0953-4075/43/10/101002} {\bibfield  {journal} {\bibinfo  {journal}
  {J. Phys. B: At. Mol. Opt. Phys.}\ }\textbf {\bibinfo {volume} {43}},\
  \bibinfo {pages} {101002} (\bibinfo {year} {2010})}\BibitemShut {NoStop}%
\bibitem [{\citenamefont {Pollack}\ \emph {et~al.}(2009)\citenamefont
  {Pollack}, \citenamefont {Dries},\ and\ \citenamefont
  {Hulet}}]{pollack:2009_science}%
  \BibitemOpen
  \bibfield  {author} {\bibinfo {author} {\bibfnamefont {S.~E.}\ \bibnamefont
  {Pollack}}, \bibinfo {author} {\bibfnamefont {D.}~\bibnamefont {Dries}}, \
  and\ \bibinfo {author} {\bibfnamefont {R.~G.}\ \bibnamefont {Hulet}},\
  }\bibfield  {title} {\enquote {\bibinfo {title} {Universality in three- and
  {Four-Body} bound states of ultracold atoms},}\ }\href {\doibase
  10.1126/science.1182840} {\bibfield  {journal} {\bibinfo  {journal}
  {Science}\ }\textbf {\bibinfo {volume} {326}},\ \bibinfo {pages} {1683--1685}
  (\bibinfo {year} {2009})}\BibitemShut {NoStop}%
\bibitem [{zac()}]{zaccanti:2009_natphys}%
  \BibitemOpen
  \bibfield  {title} {\enquote {\bibinfo {title} {Observation of an efimov
  spectrum in an atomic system},}\ }\href@noop {} {\ }\BibitemShut {NoStop}%
\bibitem [{\citenamefont {Ferlaino}\ \emph {et~al.}(2009)\citenamefont
  {Ferlaino}, \citenamefont {Knoop}, \citenamefont {Berninger}, \citenamefont
  {Harm}, \citenamefont {{D’Incao}}, \citenamefont {Nägerl},\ and\
  \citenamefont {Grimm}}]{ferlaino:2009_phys.rev.lett.}%
  \BibitemOpen
  \bibfield  {author} {\bibinfo {author} {\bibfnamefont {F.}~\bibnamefont
  {Ferlaino}}, \bibinfo {author} {\bibfnamefont {S.}~\bibnamefont {Knoop}},
  \bibinfo {author} {\bibfnamefont {M.}~\bibnamefont {Berninger}}, \bibinfo
  {author} {\bibfnamefont {W.}~\bibnamefont {Harm}}, \bibinfo {author}
  {\bibfnamefont {J.~P.}\ \bibnamefont {{D’Incao}}}, \bibinfo {author}
  {\bibfnamefont {{H.-C.}}\ \bibnamefont {Nägerl}}, \ and\ \bibinfo {author}
  {\bibfnamefont {R.}~\bibnamefont {Grimm}},\ }\bibfield  {title} {\enquote
  {\bibinfo {title} {Evidence for universal {Four-Body} states tied to an
  efimov trimer},}\ }\href {\doibase 10.1103/PhysRevLett.102.140401} {\bibfield
   {journal} {\bibinfo  {journal} {Phys. Rev. Lett.}\ }\textbf {\bibinfo
  {volume} {102}},\ \bibinfo {pages} {140401} (\bibinfo {year}
  {2009})}\BibitemShut {NoStop}%
\bibitem [{\citenamefont {Zenesini}\ \emph {et~al.}(2012)\citenamefont
  {Zenesini}, \citenamefont {Huang}, \citenamefont {Berninger}, \citenamefont
  {Besler}, \citenamefont {Nägerl}, \citenamefont {Ferlaino}, \citenamefont
  {Grimm}, \citenamefont {Greene},\ and\ \citenamefont {von
  Stecher}}]{zenesini:2012_}%
  \BibitemOpen
  \bibfield  {author} {\bibinfo {author} {\bibfnamefont {A.}~\bibnamefont
  {Zenesini}}, \bibinfo {author} {\bibfnamefont {B.}~\bibnamefont {Huang}},
  \bibinfo {author} {\bibfnamefont {M.}~\bibnamefont {Berninger}}, \bibinfo
  {author} {\bibfnamefont {S.}~\bibnamefont {Besler}}, \bibinfo {author}
  {\bibfnamefont {H.~{-C}}\ \bibnamefont {Nägerl}}, \bibinfo {author}
  {\bibfnamefont {F.}~\bibnamefont {Ferlaino}}, \bibinfo {author}
  {\bibfnamefont {R.}~\bibnamefont {Grimm}}, \bibinfo {author} {\bibfnamefont
  {Chris~H}\ \bibnamefont {Greene}}, \ and\ \bibinfo {author} {\bibfnamefont
  {J.}~\bibnamefont {von Stecher}},\ }\bibfield  {title} {\enquote {\bibinfo
  {title} {Resonant {Five-Body} recombination in an ultracold gas},}\ }\href
  {http://arxiv.org/abs/1205.1921} {\  (\bibinfo {year} {2012})},\ \bibinfo
  {note} {{arXiv:1205.1921} [cond-mat.quant-gas]}\BibitemShut {NoStop}%
\bibitem [{\citenamefont {Esry}\ \emph {et~al.}(1996)\citenamefont {Esry},
  \citenamefont {Lin},\ and\ \citenamefont {Greene}}]{esry:1996_phys.rev.a}%
  \BibitemOpen
  \bibfield  {author} {\bibinfo {author} {\bibfnamefont {B.}~\bibnamefont
  {Esry}}, \bibinfo {author} {\bibfnamefont {C.}~\bibnamefont {Lin}}, \ and\
  \bibinfo {author} {\bibfnamefont {Chris}\ \bibnamefont {Greene}},\ }\bibfield
   {title} {\enquote {\bibinfo {title} {Adiabatic hyperspherical study of the
  helium trimer},}\ }\href {\doibase 10.1103/PhysRevA.54.394} {\bibfield
  {journal} {\bibinfo  {journal} {Phys. Rev. A}\ }\textbf {\bibinfo {volume}
  {54}},\ \bibinfo {pages} {394--401} (\bibinfo {year} {1996})}\BibitemShut
  {NoStop}%
\bibitem [{\citenamefont {Kievsky}\ \emph {et~al.}(1997)\citenamefont
  {Kievsky}, \citenamefont {Marcucci}, \citenamefont {Rosati},\ and\
  \citenamefont {Viviani}}]{kievsky:1997_few-bodysyst}%
  \BibitemOpen
  \bibfield  {author} {\bibinfo {author} {\bibfnamefont {A.}~\bibnamefont
  {Kievsky}}, \bibinfo {author} {\bibfnamefont {L.~E.}\ \bibnamefont
  {Marcucci}}, \bibinfo {author} {\bibfnamefont {S.}~\bibnamefont {Rosati}}, \
  and\ \bibinfo {author} {\bibfnamefont {M.}~\bibnamefont {Viviani}},\
  }\bibfield  {title} {\enquote {\bibinfo {title} {{High-Precision} calculation
  of the triton ground state within the {Hyperspherical-Harmonics} method},}\
  }\href {\doibase 10.1007/s006010050049} {\bibfield  {journal} {\bibinfo
  {journal} {{Few-Body} Syst}\ }\textbf {\bibinfo {volume} {22}},\ \bibinfo
  {pages} {1--10} (\bibinfo {year} {1997})}\BibitemShut {NoStop}%
\bibitem [{\citenamefont {Naidon}\ \emph {et~al.}(2012)\citenamefont {Naidon},
  \citenamefont {Hiyama},\ and\ \citenamefont {Ueda}}]{naidon:2012_phys.rev.a}%
  \BibitemOpen
  \bibfield  {author} {\bibinfo {author} {\bibfnamefont {Pascal}\ \bibnamefont
  {Naidon}}, \bibinfo {author} {\bibfnamefont {Emiko}\ \bibnamefont {Hiyama}},
  \ and\ \bibinfo {author} {\bibfnamefont {Masahito}\ \bibnamefont {Ueda}},\
  }\bibfield  {title} {\enquote {\bibinfo {title} {Universality and the
  three-body parameter of {{\textasciicircum}{4}He} trimers},}\ }\href
  {\doibase 10.1103/PhysRevA.86.012502} {\bibfield  {journal} {\bibinfo
  {journal} {Phys. Rev. A}\ }\textbf {\bibinfo {volume} {86}},\ \bibinfo
  {pages} {012502} (\bibinfo {year} {2012})}\BibitemShut {NoStop}%
\bibitem [{\citenamefont {Berninger}\ \emph {et~al.}(2011)\citenamefont
  {Berninger}, \citenamefont {Zenesini}, \citenamefont {Huang}, \citenamefont
  {Harm}, \citenamefont {Nägerl}, \citenamefont {Ferlaino}, \citenamefont
  {Grimm}, \citenamefont {Julienne},\ and\ \citenamefont
  {Hutson}}]{berninger:2011_phys.rev.lett.}%
  \BibitemOpen
  \bibfield  {author} {\bibinfo {author} {\bibfnamefont {M.}~\bibnamefont
  {Berninger}}, \bibinfo {author} {\bibfnamefont {A.}~\bibnamefont {Zenesini}},
  \bibinfo {author} {\bibfnamefont {B.}~\bibnamefont {Huang}}, \bibinfo
  {author} {\bibfnamefont {W.}~\bibnamefont {Harm}}, \bibinfo {author}
  {\bibfnamefont {{H.-C.}}\ \bibnamefont {Nägerl}}, \bibinfo {author}
  {\bibfnamefont {F.}~\bibnamefont {Ferlaino}}, \bibinfo {author}
  {\bibfnamefont {R.}~\bibnamefont {Grimm}}, \bibinfo {author} {\bibfnamefont
  {P.}~\bibnamefont {Julienne}}, \ and\ \bibinfo {author} {\bibfnamefont
  {J.}~\bibnamefont {Hutson}},\ }\bibfield  {title} {\enquote {\bibinfo {title}
  {Universality of the {Three-Body} parameter for efimov states in ultracold
  cesium},}\ }\href {\doibase 10.1103/PhysRevLett.107.120401} {\bibfield
  {journal} {\bibinfo  {journal} {Phys. Rev. Lett.}\ }\textbf {\bibinfo
  {volume} {107}},\ \bibinfo {pages} {120401} (\bibinfo {year}
  {2011})}\BibitemShut {NoStop}%
\bibitem [{\citenamefont {Frederico}\ \emph {et~al.}(1999)\citenamefont
  {Frederico}, \citenamefont {Tomio}, \citenamefont {Delfino},\ and\
  \citenamefont {Amorim}}]{frederico:1999_phys.rev.a}%
  \BibitemOpen
  \bibfield  {author} {\bibinfo {author} {\bibfnamefont {T.}~\bibnamefont
  {Frederico}}, \bibinfo {author} {\bibfnamefont {Lauro}\ \bibnamefont
  {Tomio}}, \bibinfo {author} {\bibfnamefont {A.}~\bibnamefont {Delfino}}, \
  and\ \bibinfo {author} {\bibfnamefont {A.~E.~A.}\ \bibnamefont {Amorim}},\
  }\bibfield  {title} {\enquote {\bibinfo {title} {Scaling limit of weakly
  bound triatomic states},}\ }\href {\doibase 10.1103/PhysRevA.60.R9}
  {\bibfield  {journal} {\bibinfo  {journal} {Phys. Rev. A}\ }\textbf {\bibinfo
  {volume} {60}},\ \bibinfo {pages} {R9--R12} (\bibinfo {year}
  {1999})}\BibitemShut {NoStop}%
\bibitem [{\citenamefont {Gattobigio}\ \emph
  {et~al.}(2011{\natexlab{c}})\citenamefont {Gattobigio}, \citenamefont
  {Kievsky},\ and\ \citenamefont
  {Viviani}}]{gattobigio:2011_j.phys.:conf.ser.}%
  \BibitemOpen
  \bibfield  {author} {\bibinfo {author} {\bibfnamefont {M}~\bibnamefont
  {Gattobigio}}, \bibinfo {author} {\bibfnamefont {A}~\bibnamefont {Kievsky}},
  \ and\ \bibinfo {author} {\bibfnamefont {M}~\bibnamefont {Viviani}},\
  }\bibfield  {title} {\enquote {\bibinfo {title} {Few-nucleon bound states
  using the unsymmetrized {HH} expansion},}\ }\href {\doibase
  10.1088/1742-6596/336/1/012006} {\bibfield  {journal} {\bibinfo  {journal}
  {J. Phys.: Conf. Ser.}\ }\textbf {\bibinfo {volume} {336}},\ \bibinfo {pages}
  {012006} (\bibinfo {year} {2011}{\natexlab{c}})}\BibitemShut {NoStop}%
\bibitem [{\citenamefont {Hadizadeh}\ \emph {et~al.}(2011)\citenamefont
  {Hadizadeh}, \citenamefont {Yamashita}, \citenamefont {Tomio}, \citenamefont
  {Delfino},\ and\ \citenamefont {Frederico}}]{hadizadeh:2011_phys.rev.lett.}%
  \BibitemOpen
  \bibfield  {author} {\bibinfo {author} {\bibfnamefont {M.~R.}\ \bibnamefont
  {Hadizadeh}}, \bibinfo {author} {\bibfnamefont {M.~T.}\ \bibnamefont
  {Yamashita}}, \bibinfo {author} {\bibfnamefont {Lauro}\ \bibnamefont
  {Tomio}}, \bibinfo {author} {\bibfnamefont {A.}~\bibnamefont {Delfino}}, \
  and\ \bibinfo {author} {\bibfnamefont {T.}~\bibnamefont {Frederico}},\
  }\bibfield  {title} {\enquote {\bibinfo {title} {Scaling properties of
  universal tetramers},}\ }\href {\doibase 10.1103/PhysRevLett.107.135304}
  {\bibfield  {journal} {\bibinfo  {journal} {Phys. Rev. Lett.}\ }\textbf
  {\bibinfo {volume} {107}},\ \bibinfo {pages} {135304} (\bibinfo {year}
  {2011})}\BibitemShut {NoStop}%
\bibitem [{\citenamefont {Deltuva}(2012)}]{deltuva:2012_}%
  \BibitemOpen
  \bibfield  {author} {\bibinfo {author} {\bibfnamefont {A.}~\bibnamefont
  {Deltuva}},\ }\bibfield  {title} {\enquote {\bibinfo {title} {Properties of
  universal bosonic tetramers},}\ }\href {http://arxiv.org/abs/1202.0167} {\
  (\bibinfo {year} {2012})},\ \bibinfo {note} {{arXiv:1202.0167}
  [physics.atom-ph]}\BibitemShut {NoStop}%
\bibitem [{\citenamefont {von
  Stecher}(2011)}]{von_stecher:2011_phys.rev.lett.}%
  \BibitemOpen
  \bibfield  {author} {\bibinfo {author} {\bibfnamefont {Javier}\ \bibnamefont
  {von Stecher}},\ }\bibfield  {title} {\enquote {\bibinfo {title} {Five- and
  {Six-Body} resonances tied to an efimov trimer},}\ }\href {\doibase
  10.1103/PhysRevLett.107.200402} {\bibfield  {journal} {\bibinfo  {journal}
  {Phys. Rev. Lett.}\ }\textbf {\bibinfo {volume} {107}},\ \bibinfo {pages}
  {200402} (\bibinfo {year} {2011})}\BibitemShut {NoStop}%
\end{thebibliography}
\end{document}